\documentclass{aa}  
\usepackage{graphicx}
\usepackage{txfonts}
\usepackage[colorlinks=true,allcolors=blue]{hyperref}
\usepackage{color}
\usepackage{ulem}

\newcommand{\Ldet}{$\mathcal{L}_{\rm DET}$}
\newcommand{\Lext}{$\mathcal{L}_{\rm EXT}$}

\begin{document} 

\title{Offset between X-ray and optical centers in clusters of galaxies: connecting eROSITA data and simulations}
\author{
R. Seppi\inst{1}\thanks{E-mail: rseppi@mpe.mpg.de} \and
J. Comparat\inst{1} \and
K. Nandra\inst{1} \and
K. Dolag\inst{3} \and
V. Biffi\inst{2} \and
E. Bulbul\inst{1} \and
A. Liu\inst{1} \and
V. Ghirardini\inst{1} \and
J. Ider-Chitham\inst{1}
}
\institute{
Max-Planck-Institut f\"{u}r extraterrestrische Physik (MPE), Giessenbachstraße 1, D-85748 Garching bei M\"unchen, Germany
\and
Physics Department, Astronomy Unit, Trieste University, Trieste, Italy; Observatory of Trieste, INAF, Trieste, Italy
\and
Universit\"ats-Sternwarte, Fakult\"at f\"ur Physik, Ludwig-Maximilians-Universit\"at M\"unchen, Scheinerstr.1, 81679 M\"unchen, Germany
}

\date{Accepted XXX. Received YYY; in original form ZZZ}

\abstract{
The characterization of the dynamical state of galaxy clusters is key to study their evolution, evaluate their selection, and use them as a cosmological probe. In this context, the offsets between different definitions of the center have been used to estimate the cluster disturbance. }
{
%
Our goal is to study the distribution of the offset between the X-ray and optical centers in clusters of galaxies. We study the offset for clusters detected by the extended ROentgen Survey with an Imaging Telescope Array (eROSITA) on board the Spectrum-Roentgen-Gamma
(SRG) observatory. We aim to connect observations to predictions by hydrodynamical simulations and N-body models. We assess the astrophysical effects affecting the displacements.}
{
%
We measure the offset for clusters observed in the eROSITA Final Equatorial-Depth Survey (eFEDS) and the first eROSITA all-sky survey (eRASS1). We focus on a subsample of 87 massive eFEDS clusters at low redshift, with M$_{\rm 500c}$>1$\times$10$^{14}$ M$_\odot$ and 0.15<z<0.4. We compare the displacements in such sample to the ones predicted by the TNG and the Magneticum simulations. We additionally link the observations to the offset parameter X$_{\rm off}$ measured on dark matter halos in N-body simulations, using the hydrodynamical simulations as a bridge.}
{%
We find that on average the eFEDS clusters show a smaller offset compared to eRASS1, because the latter contains a larger fraction of massive and disturbed structures. We measure an average offset of $\Delta_{\rm X-O}$=76.3$_{\rm -27.1}^{\rm +30.1}$ kpc, when focusing on the subsample of 87 eFEDS clusters. This is in agreement with the predictions from TNG and Magneticum, and the distribution of X$_{\rm off}$ from dark matter only (DMO) simulations. However, the tails of the distributions are different. Using $\Delta_{\rm X-O}$ to classify relaxed and disturbed clusters, we measure a relaxed fraction of 31$\%$ in the eFEDS subsample.  Finally, we find a correlation between the offset measured on hydrodynamical simulations and X$_{\rm off}$ measured on their parent dark matter-only run and calibrate a relation between them. 
}
{
We conclude that there is good agreement between the offsets measured in eROSITA data and the predictions from simulations. Baryonic effects cause a decrement (increment) in the low (high) offset regime compared to the X$_{\rm off}$ distribution from dark matter-only simulations. The offset--X$_{\rm off}$ relation provides an accurate prediction of the true X$_{\rm off}$ distribution in Magneticum and TNG. It allows introducing the offsets in a cosmological context with a new method, in order to marginalize on selection effects related to the cluster dynamical state.
}

\maketitle

\section{Introduction}

Clusters of galaxies are the most massive virialized structures in the Universe. They represent the final step of the hierarchical process of structure formation, where lower mass objects merge to form bigger structures. Their growth through cosmic time strongly depends on the evolution history of the Universe, making them a powerful cosmological probe \citep{Allen2011, weinberg_2013_review, Mantz2015cosmology, Clerc2022arXiv220311906C_review}.\\
Galaxy clusters are identified in different wavelengths, given their distinctive observational features, such as an over-density of red sequence galaxies \citep{Gladders2005ApJ, Yang2007ApJ}; a distortion of the images of background galaxies by strong and weak gravitational lensing \citep{Maturi2005clusters_WL, Miyazaki2018shearsel}; the X-ray emission due to thermal bremsstrahlung from the hot intra-cluster gas \citep{Ebeling1998MNRAS, Bohringer2000ApJ, Vikhlinin2009ApJ...692.1060V, Pierre2016XLL}; and the distortion of the cosmic microwave background (CMB) spectrum due to the Sunyaev-Zel'dovich effect \citep{Sunyaev1972CoASP...4..173S, Planck2014A&A...571A..20P}.

The definition and identification of the cluster center is a key step in their analysis. From a purely dark matter standpoint, it is natural to consider the deepest point in the potential well of the dark matter halo hosting the cluster.
This is traced best by lensing observations \citep{Zitrin2012MNRAS}. Other possibilities involve the peak or the centroid of the gas emission in the X-ray and millimeter bands \citep[][]{Rossetti2016dyn_state, Gupta2017_pressureprof_magneticum}. Finally, a cluster center is also identified using optical and infrared data \citep[][]{Ota2020miscentering}, by considering the position of the central galaxy (CG), for example, the brightest cluster galaxy (BCG). \\
An agreement between these definitions is expected if the dark matter halo and different baryonic components are completely relaxed and in equilibrium within the potential well of the cluster. However, galaxy clusters are rarely in complete dynamical equilibrium. They assembled at late times, undergoing mergers. This leads to disturbed mass distribution and an offset between different definitions of the cluster center.

A deeper insight into this topic is now possible thanks to the extended ROentgen Survey with an Imaging Telescope Array \citep[eROSITA, ][]{Merloni2012, Predehl2021A&Aerosita} onboard Spectrum-Roentgen-Gamma (SRG). It is a collaboration between the German and the Russian consortia. The full sky is split in half between them. This X-ray instrument uses seven telescope modules with 54 nested mirror shells each. Its point spread function (PSF) has a half energy width (HEW) of about 15\arcsec\ for each module. As a part of the calibration and performance verification phase, SRG-eROSITA performed a mini-survey of the 140 square degrees eROSITA
Final Equatorial Depth Survey \citep[eFEDS,][]{Brunner2022_efedscat}, to test its capabilities. SRG-eROSITA was launched in July 2019 and finished its first all-sky survey (eRASS1) in June 2020. This X-ray telescope is scheduled to complete eight all-sky scans, providing X-ray observations for $\sim$100\,000 clusters, reaching unmatched data quality for the X-ray view of these objects and their astrophysical and cosmological purposes. The goal of this article is to exploit the optical follow-up of eROSITA clusters to grasp a deeper knowledge of their offset between X-ray and optical centers, and compare the result to predictions from hydrodynamical simulations and N-body models. We summarize the key aspects of various definitions of the cluster center, and their implications in terms of the dynamical state. 

Historically, it is often assumed that the BCG is the central galaxy of the cluster, i.e. the one closest to the deepest point of the halo potential well \citep{vandenBosch2004MNRAS.352.1302V, Weinmann2006MNRAS.366....2W}. Therefore, the BCG is used to define the cluster center in the optical band. However, recent works show that this is not always the case \citep{Einasto2011ApJ, Lange2018bcg}. In particular, \cite{Skibba2011BCG} find that the BCG is not the central galaxy in $\sim$25$\%$ of galaxy group-like halos. This fraction increases to $\sim$45$\%$ for clusters of galaxies. The development of new techniques to identify clusters in the optical band reduced the mis-centered fraction. The red-sequence Matched-filter Probabilistic Percolation \citep[redMaPPer,][]{Rykoff2014redmapper, Rykoff2016ApJS..224....1R} is a cluster finding algorithm for photometric surveys, such as the Sloan Digital Sky Survey \citep[SDSS,][]{York2000sloan}, the Dark Energy Survey \citep[DES,][]{DES2005astro.ph.10346T}, and the Large Synoptic Survey Telescope (LSST) at the Rubin Observatory \citep{2009lsst}. It allows locating the cluster optical center with additional information such as redshift and local galaxy density. \cite{Hoshino2015MNRAS.452..998H} analyzed the occupation of luminous red galaxies with redMaPPer centering probabilities and show that the BCG is not the central galaxy in 20-30$\%$ of the clusters. The centering algorithm of redMaPPer is based on assigning a probability to each member of being the central galaxy and provides a more consistent definition of the optical center. \\
\cite{Rozo2014redmapper_xray_sz} studied the performance of redMaPPer on SDSS data by comparing the optical catalog to overlapping X-ray and SZ data. They find that about 80$\%$ of the clusters are well centered, with offsets smaller than 50 kpc. The remaining 20$\%$ consists of mergers, which exhibit much larger offsets even up to 300 kpc.
The displacement decreases at low redshift \citep{Gozaliasl2019MNRAS.483.3545G}. This is in agreement with the hierarchical scenario, where structures relax at late times.
The offset between peaks in various bands has been exploited to identify relaxed and disturbed systems \citep[][]{Mann2012MNRAS.420.2120M_offset, Rossetti2016dyn_state, Rossetti2017MNRAS, Oguri2018PASJ...70S..20O, Ota2020miscentering, Ota2022_offset}.\\
A detailed description of these offsets and their link to the cluster dynamical state is also important to assess possible biases and selection effects, especially in the current era of precision cosmology. For instance, a partial knowledge of the baryon physics affecting the evolution of galaxy clusters biases scaling relations between observables and clusters masses \citep[][]{Bahar2022A&A...661A...7B, Chiu2022A&Aefeds}, which ultimately impact cosmological results \citep{Chisari2019OJApsims, Genel2019ApJ...871...21G, Salvati2020, Debackere2021MNRAS.505..593D, Castro2021MNRAS.500.2316C}. The disturbance and morphological diversity of these extended objects make the understanding of selection effects non-trivial \citep{Weissmann2013_morphology, Cao2020morphology}. In addition, baryonic properties potentially affect the selection of clusters in astronomical surveys. They might alter the values of a specific observable, which ends up affecting the number of objects in the sample compared to an unbiased theoretical prediction. \\
X-ray observations of galaxy clusters suffer from the cool core bias \citep{Eckert2011, Kafer2019}. The largest dark matter halos hosting massive clusters of galaxies assemble at late times. Some clusters did not have enough time to dynamically relax and develop a cool core, which would be detected as a peak in the X-ray surface brightness profile. This feature biases the detection towards relaxed structures with a cool core, affecting the completeness of X-ray-selected samples of galaxy clusters. At fixed mass and redshift, cool core clusters are therefore more probable to be detected compared to non-cool core ones. The cool core bias is expected to play a role in the characterization of clusters as extended sources. Cool core clusters possibly have a higher probability of being confused for point sources, because the peaked emissivity in the central region dominates over the extended emission in the cluster outskirts \citep{Somboonpanyakul2021ApJ...910...60S, Bulbul2022A&A_cluindisguise}. This has an impact on cosmological studies using the halo mass function \citep{Seppi2021A&A...652A.155S}. Therefore, it is necessary to take this selection effect into account. The evidence of the cool core bias in X-ray-selected samples has been highlighted by different works, especially when comparing X-ray to SZ selected samples, which are not affected by such bias due to the lower sensibility to the central gas density \citep[][]{Hudson2010, Eckert2011, Rossetti2017MNRAS, Andrade-Santos2017ApJ...843...76A, Lovisari2017ApJ...846...51L, Campitiello2022arXiv220511326G_checkmate}. 
However, other studies do not find a significant preference for relaxed clusters \citep[e.g.][]{Mantz2015MNRAS.449..199M, Nurgaliev2017ApJ...841....5N, McDonald2017ApJ...843...28M, DeLuca2020300_dyn_state}. This topic has been analyzed with eROSITA data by \citet{Ghirardini2022A&Amorph}, who did not find a clear bias towards relaxed structures. In addition, \citet{Bulbul2022A&A_cluindisguise} find a preference for cool cores only when looking for clusters cataloged as point sources. Strong evidence for the cool core bias in the point-like sample is also predicted by eROSITA simulations \citep[][]{Seppi2022A&A_eRASS1sim}. \\
The fraction of mass in substructures, central entropy, spin, and offset parameters give additional insight into the dynamical state \citep{Meneghetti2014ApJ...797...34M, Biffi2016HE, Henson2016, DeLuca2020300_dyn_state, Seppi2021A&A...652A.155S}. \\
A precise knowledge of the cluster center would benefit various studies, such as the measure of weak lensing profiles \citep[][]{Chiu2022A&Aefeds}, where the error in the measurement may be reduced with a better comprehension of the miscentering \citep[][]{George2012centering, Zhang2019MNRAS.487.2578Z, Yan2020MNRAS_bahamas, Ota2022_offset}; or detailed comparison of cluster density profiles with simulations \citep[][]{Zhuravleva2013MNRAS.428.3274Z, Diemer2022MNRAS.513..573D}.

We measure the offset between the position of the X-ray and the optical centers for eROSITA clusters. We use two samples: eFEDS \citep[][]{2022A&A_LiuAng_eFEDS_clu}, and eRASS1. The optical follow-up is performed with redMaPPer, making use of the prior knowledge of the X-ray position \citep[][]{IderChitham2020MNRAS.499.4768I}. We study the distribution of the offsets and different physical effects affecting them.
We look for a link between observations and the dynamical state of dark matter halos in N-body simulations \citep{Klypin2016, Seppi2021A&A...652A.155S}. We consider the offset parameter ($X_{\rm off}$), that is the displacement between the peak of the mass profile and the center of mass of dark matter halos. \citet{Seppi2021A&A...652A.155S} calibrated a mass function model that allows marginalizing on variables related to the dynamical state. Instead, we marginalize on mass, predict the distribution of $X_{\rm off}$, and compare it to the displacement between X-ray and optical centers. We exploit hydrodynamical simulations to develop this connection. We use the Magneticum \citep{Biffi2013MNRAS, Hirschmann2014MNRAS.442.2304H, Biffi2018lightcone, Ragagnin2017webportal} and the Illustris-TNG \citep{Pillepich2018MNRASTNG_method, Nelson2019TNG} simulations.

We summarize the eROSITA data, its processing, the optical follow-up, and the hydrodynamical simulations in Sect. \ref{sec:data}. We describe our method for computing the offsets in eROSITA data, in simulations, and using the N-body model from \citet{Seppi2021A&A...652A.155S} in Sect. \ref{sec:method}. We present the distributions of the offsets and our results in Sect. \ref{sec:results}. We discuss our findings and how to use the offsets in a cosmological framework in Sect. \ref{sec:discussion}. We finally summarize our results in Sect. \ref{sec:conclusions}.

\section{Data}
\label{sec:data}
In this Section, we describe the X-ray observations, the optical follow-up, and the hydrodynamical simulations used in this work.

\subsection{eROSITA}

We use X-ray data from the eROSITA X-ray telescope. The observations are processed with the eROSITA Standard Analysis Software System \citep[eSASS,][]{Brunner2022_efedscat}. The detection focuses on the soft X-ray band (0.2--2.3 keV) and relies on a modified sliding box algorithm (\texttt{erbox}). It marks potential sources that emerge over the background by a chosen confidence threshold. Such regions are then masked and the remaining source-free map is used to create a background map by the \texttt{erbackmap} tool. The combination of these two algorithms is run twice to obtain a more reliable background map. Finally, each box marked as a potential source is fitted by the maximum likelihood algorithm \texttt{ermldet}. It uses PSF-fitting to measure the source parameters, such as position, count rate, detection likelihood (DET\_LIKE or \Ldet), extent likelihood (EXT\_LIKE or \Lext) and the source extent, equal to the best-fitting core radius of the $\beta$-model \citep[][]{CavaliereFuscoFermiano1976A&A....49..137C}. DET\_LIKE is related to the probability of the source being a background fluctuation, and EXT\_LIKE is the likelihood of the $\beta$-model over the point-like model convolved with the PSF. A detailed discussion on cluster detection with eROSITA is given by \citet{Seppi2022A&A_eRASS1sim}. In addition, we study the probability of membership for all galaxy members in each cluster using redMaPPer \citep[][]{Rykoff2014redmapper, IderChitham2020MNRAS.499.4768I} in scanning mode, making use of the prior knowledge of the X-ray position.\\
Given the relatively small area of eFEDS and the shallow depth of eRASS1, an accurate description of the high-z cluster population is not feasible.

\subsubsection{eFEDS}
\label{subsubsec:eFEDS_data}
During the Calibration and Performance Verification Phase (CalPV), the eROSITA Final Equatorial Depth Survey (eFEDS) has been carried out. eFEDS was designed with the goal of verifying the survey capabilities of eROSITA. This mini-survey covers an area of $\sim$140 deg$^2$ in the equatorial region (126$^\circ$ < RA< 146$^\circ$, -3$^\circ$ < DEC < +6$^\circ$). It was covered with a vignetted (unvignetted) exposure time of $\sim$1.2 ks ($\sim$2.2 ks), a similar value compared to the final all-sky survey (eRASS:8) in the equatorial region.\\
We use the cluster catalog from \citet{2022A&A_LiuAng_eFEDS_clu}. It includes 542 clusters with \Ldet>5 and \Lext>6. The clusters are confirmed in the optical band and the redshifts are measured with the Multi-Component Matched Filter (MCMF) cluster confirmation tool \citep{Klein2018Mcmf, Klein2022A&A.eFEDS}, combining optical data from different surveys such as the Hyper
Suprime-Cam (HSC) Subaru Strategic Program \citep[HSC-SSP,][]{Oguri2018PASJ...70S..20O}, the Dark Energy Camera Legacy Survey \citep[DECaLS,][]{Dey2019AJDESI}), the Sloan Digital Sky Survey \citep[SDSS,][]{Blanton2017AJsdss4}, the 2MASS Redshift Survey \citep[2MRS,][]{Huchra2012ApJS..199...26H}, and the Galaxy And Mass Assembly \citep[GAMA,][]{Driver2011}. A detailed weak-lensing study on HSC observations by \citet{Chiu2022A&Aefeds} provides halo masses for a subsample of 434 eFEDS clusters. 

\subsubsection{eRASS1}
\label{subsubsec:eRASS1_data}
eROSITA performed its first scan of the whole sky during the first six months of the survey phase, from December 13th 2019 until June 11th 2020, completing the first all-sky survey (eRASS1\footnote{\url{https://www.mpe.mpg.de/7461950/erass1-presskit}}). Given the scanning strategy of the telescope, the exposure time depends on the angular position on the sky. Shallow regions around the ecliptic equator are covered for less than 100 seconds, while deep areas around the ecliptic poles are observed for more than 1.2 ks \citep[see][for more details]{Predehl2021A&Aerosita}. The average exposure time of eRASS1 is $\sim$250 s. We use the German half of the sky (eROSITA\_DE). The majority of the area overlaps with different optical surveys, such as the Dark Energy Camera Legacy Survey \citep[DECaLS,][]{Dey2019AJDESI}, the Dark Energy Survey \citep[DES,][]{Sevilla-Noarbe2021ApJSDES_catalog}, and the Kilo-Degree Survey \citep[KiDS,][]{Kuijken2019A&AKiDS_dr4}. The measure of redshifts and optical properties is carried out by redMaPPer \citep[][]{Rykoff2014redmapper, IderChitham2020MNRAS.499.4768I}.\\
We focus on the X-ray position measured by eSASS, the optical centers, and the redshift provided by redMaPPer. The scaling relation between X-ray luminosity and mass from \citet{Chiu2022A&Aefeds} provides a mass estimate for eRASS1 clusters.

\subsection{Simulations}
In this work, we use the Illustris-TNG and the Magneticum simulations. The main numerical and cosmological parameters for the two simulations are written in Table \ref{tab:simulation_parameters}.

\begin{table}
    \caption{Numerical and physical parameters describing the Magneticum and Illustris-TNG simulations.}
    \centering
    \begin{tabular}{|c|c|c|}
    \hline
    \rule{0pt}{2.2ex} & \textbf{Magneticum-Box2/hr} & \textbf{TNG-300-1} \\
    \hline
    \rule{0pt}{2.2ex} Box size [Mpc/\textit{h}] & 352 & 205 \\
    $\Omega_{\rm M}$ & 0.272 & 0.3089 \\
    $\Omega_{\rm B}$ & 0.0456 & 0.0486 \\
    $\Omega_{\rm \Lambda}$ & 0.728 & 0.6911 \\
    $\sigma_{\rm 8}$ & 0.809 & 0.8159 \\
    H$_{\rm 0}$ & 70.4 & 67.74 \\
    n$_{\rm s}$ & 0.963 & 0.9667 \\
    N particles & 2$\times$ 1584$^3$ & 2500$^3$ \\
    M$_{\rm DM}$ [M$_\odot$/\textit{h}] & 6.9$\times$10$^8$ & 5.9$\times$10$^7$ \\
    \hline
    \end{tabular}
    \footnotesize{\textbf{Notes.} Volume: total comoving volume covered by the simulation, $\Omega_{\rm M}$: total matter density parameter, $\Omega_{\rm B}$: baryonic matter density parameter, $\Omega_{\rm \Lambda}$: dark energy density parameter, $\sigma_{\rm 8}$: normalization of the linear matter power spectrum, $H_{\rm 0}$: Hubble constant, n$_{\rm s}$: initial slope of the linear matter power spectrum, N particles: total number of dark matter particles in the simulation, M$_{\rm DM}$: mass of the dark matter particles.}
    \label{tab:simulation_parameters}
\end{table}

\subsubsection{Magneticum}
The Magneticum simulation suite\footnote{\url{http://www.magneticum.org}} is a set of cosmological hydrodynamical and dark-matter-only simulations \citep[][]{Biffi2013MNRAS, Hirschmann2014MNRAS.442.2304H, Dolag2015IAUGA..2250156D, Steinborn2015MNRAS.448.1504S, Ragagnin2017webportal, Dolag2017_Magneticum_metals, Singh2020MNRAS.494.3728S}, spanning different ranges of resolution and box size. These simulations are run with the TreePM-SPH code \texttt{P-GADGET3} \citep[][]{Springel2005}. Multiple processes regulated by baryonic physics are taken into account in the simulation, such as radiative cooling \citep[][]{Wiersma2009MNRAS.399..574W}, heating due to star formation, supernovae, galactic winds \citep[][]{Springel2003MNRAS_SF}, chemical enrichment \citep[][]{Tornatore2007MNRAS.382.1050T}, and AGN feedback processes \citep[][]{Fabjan2010MNRAS_AGNfeedback}. The Magneticum simulations are successful at reproducing the black hole mass density \citep[][]{DiMatteo2008ApJ...676...33D}, the AGN luminosity function \citep[][]{Hirschmann2014MNRAS.442.2304H, Steinborn2016MNRAS_AGN, Biffi2018lightcone}, morphological properties of galaxies \citep[][]{Teklu2015ApJ...812...29T, Remus2017MNRAS_Magneticum}, and the pressure profiles of galaxy clusters \citep[][]{Gupta2017_pressureprof_magneticum}. This set of simulations has been used to quantify the impact of baryons on the halo mass function \citep[][]{Bocquet2016, Castro2021MNRAS.500.2316C}, and for dedicated studies of the Large Scale Structure around merging galaxy clusters with eROSITA \citep{Biffi2022A&A_Abell339195}.\\
We focus on the \textbf{Box2/hr} simulation. It is computed assuming a WMAP cosmology \citep[][]{Komatsu2011ApJSWMAP7}. Given our interest in clusters of galaxies, it provides a great compromise between the size of the box and the resolution of the dark matter halos. The side of the simulated cube is 352 Mpc/\textit{h} (500 Mpc) large. The box contains 475 halos more massive than M$_{\rm 500c}$=1$\times$10$^{14}$ M$_\odot$ at z=0\footnote{M$_{\rm 500c}$ is the total mass of the cluster encompassed by a radius containing an average density that is 500 times larger than the critical density of the Universe.}. The resolution of the dark matter particles is 6.9$\times$10$^8$ M$_\odot$/\textit{h}, which allows measuring detailed properties of the most massive halos hosting clusters and groups. A summary of the key parameters for the simulation is reported in Table \ref{tab:simulation_parameters}.

\subsubsection{Illustris-TNG}
The Illustris-TNG project\footnote{\url{https://www.tng-project.org}} is a collection of 18 complementary hydrodynamical simulations coupled with dark-matter-only runs \citep[][]{Weinberger2017MNRAS.465.3291W, Pillepich2018MNRASTNG_method, Barnes2018, Nelson2019TNG}. It spans different box sizes, resolutions, and baryonic physics. The simulations are run with the quasi-Lagrangian code \texttt{AREPO} \citep[][]{Weinberger2020ApJSAREPO}. It includes gas radiative mechanisms, star formation, stellar evolution, supernovae explosions, the formation and accretion of supermassive black holes, and the amplification of magnetic fields. The TNG project successfully reproduces the galaxy color distribution as a function of stellar mass \citep[][]{Nelson2018MNRAS.475..624N}, the stellar mass function at recent epochs, the distribution of stellar mass inside galaxy clusters \citep[][]{Pillepich2018MNRASTNG_result}, the scaling relation between radio power and X-ray emission in galaxy clusters \citep[][]{Marinacci2018MNRAS.480.5113M}, the low redshift quasar luminosity function \citep[][]{Weinberger2018MNRAS_TNG_AGNfeedback}, the chemical evolution of gas in galaxies \citep[][]{Naiman2018MNRAS.477.1206N}, and the galaxy two-point correlation function \citep{Springel2018MNRAS.475..676S}. \\
The TNG project assumes a Planck cosmology \citep{Planck2016A&A..cosmopars}. We use the TNG-300-1 simulation. It is the largest available box, with a side of 205 Mpc/h (300 Mpc). It is smaller than Magneticum Box2/hr, and contains therefore fewer halos: 159 objects more massive than M$_{\rm 500c}$=1$\times$10$^{14}$ M$_\odot$ at z=0. However, it has a better particle resolution (see Table \ref{tab:simulation_parameters}).

\section{Method}
\label{sec:method}

In this Section, we describe how we processed and analyzed the data, and how we compared it to theoretical models and hydrodynamical simulations.

\subsection{Offset for eROSITA clusters}
\label{subsec:erosita_clusters}
We focus on eFEDS and eRASS1 clusters with EXT\_LIKE > 6. We additionally require a measure of the uncertainty on the X-ray position by eSASS (RADEC\_ERR > 0). For eRASS1, we exclude clusters that are not covered by optical surveys and are therefore lacking a measure of the optical center. We determine the X-ray center using the cluster position defined by eSASS. It is the best fit position of the PSF fit, that determines the X-ray centroid. \\
We consider two definitions of the optical center.
The first one is given by the centering algorithm of redMaPPer. It uses a Bayesian classification algorithm to locate the most probable cluster center. It is based on a local red galaxy density filter that ensures consistency between the photometric redshift of the central galaxy and the cluster. It also matches the central galaxy luminosity to an expected value given the cluster richness, which is closely related to the total number of galaxy members hosted by the cluster \citep[][]{Rykoff2014redmapper}. The optical center is not always placed on the brightest cluster galaxy. In fact, $\sim$20$\%$ of the time the central galaxy is not the brightest member \citep[][]{Rykoff2016ApJS..224....1R}. This approach provides the probability for each member to be the central cluster galaxy P$_{\rm cen}$ \citep[see Eq. 56 in][]{Rykoff2014redmapper}. 
Secondly, we consider the position of the galaxy member with the largest membership probability P$_{\rm mem}$. It is the probability that a galaxy
near a cluster is a cluster member and should not be confused with the probability of being the central galaxy P$_{\rm cen}$. It is computed for each galaxy by combining different filters. The most important one is a model of the color evolution of red-sequence galaxies as a function of redshift \citep[see Eq. 1 in][]{Rykoff2014redmapper}. \\
Given the angular positions of the X-ray and optical centers, we compute the angular offset between them and convert it to the comoving physical kiloparsec scale based on the cluster redshift, according to

\begin{equation}
    \Delta_{\rm X-O} = \frac{c}{H_{\rm 0}} \int_0^z \frac{dz}{\sqrt{\Omega_{\rm M}(1+z)^3 + \Omega_\Lambda}} \times \theta, \hspace{1.8cm} \text{[kpc]}
    \label{eq:offset}
\end{equation}
where z is the cluster redshift, c is the speed of light, H$_0$ is the Hubble constant, and $\theta$ is the angular separation between X-ray and optical positions in radians. Similarly, we measure the offset to the position of the galaxy with the largest membership probability $\Delta_{\rm X-P_{\rm mem}}$. \\
We estimate the error on the X-ray center by multiplying the uncertainty on the angular X-ray position by the physical scale per unit angle.
We estimate a systematic error on the optical center accounting for the separation between the optical center identified by redMaPPer and the position of the 5 most probable centers weighted by their centering probability, according to Equation \ref{eq:error_optcent}:

\begin{equation}
    \delta_{\rm O} = \sqrt{\sum_{\rm i=1}^{\rm N=5} (P_{\rm cen,i} \times \Delta_{\rm O-O_i})^2},
    \label{eq:error_optcent}
\end{equation}
where the index i runs on the five most probable members and $\Delta_{\rm O-O_i}$ is the offset in kpc scale between the optical center and the position of the i galaxy member. In this manner, the uncertainty $\delta_{\rm O}$ accounts for the fact that the definition of a center is complicated when there are many bright galaxies with similar probability of being the central galaxy.\\
We finally compute the cumulative distribution function (CDF) of $\Delta_{\rm X-O}$ and $\Delta_{\rm X-P_{\rm mem}}$. We do this first for the whole eRASS1 and eFEDS samples, by restricting to secure clusters with more than 20 counts and richness $\lambda$>20. These cuts discard clusters with large average errors on the X-ray position larger than 100 kpc. We obtain 182 (4564) clusters from eFEDS (eRASS1) satisfying these conditions. We then focus on a more specific sub-sample of 87 eFEDS clusters between redshift 0.15 and 0.4 and M$_{\rm 500c}$ between 1$\times$10$^{14}$ and 8$\times$10$^{14}$ M$_\odot$. The mean values of mass and reshift are M$_{\rm 500c}$=2.16$\times$10$^{14}$ M$_\odot$ and z=0.30.
We use this well-defined sample to do a comparison with simulations.

\subsection{Analytical DMO model}
\label{subsec:seppi21_model}
We propose a link between the X-ray to optical offset in observations and the theoretical model developed by \citet{Seppi2021A&A...652A.155S}. There the authors calibrated a model for the halo mass function, that additionally includes variables describing the dynamical state of dark matter halos. We are particularly interested in the offset parameter X$_{\rm off}$, that is the displacement between the center of mass of a dark matter halo and the peak of its density profile, normalized to the virial radius. Such mass function model predicts the dark matter halo abundance as a function of mass, offset parameter, and spin, offering the possibility of integrating out one or more of these variables. We marginalize on mass and spin and obtain the analytical prediction of the 1D distribution for the offset parameter \citep[see Eq. 18-20 in][]{Seppi2021A&A...652A.155S}. \\ 
We focus on the eFEDS sample, where masses have been estimated via weak gravitational lensing \citep{Chiu2022A&Aefeds}.
We compare the offsets between the X-ray and optical centers to the offset parameter in physical scales X$_{\rm off, P}$ (in kpc, i.e. not normalized to the virial radius) from N-body simulations \citep[see Eq. B.3 in][]{Seppi2021A&A...652A.155S}. We compute the halo multiplicity function dependent on the offset parameter by marginalizing on mass and detection probability according to Equation \ref{eq:fxoff}:

\begin{equation}
    f(X_{\rm off,P}) = \int_{M_{\rm low}}^{M_{\rm up}} g(\sigma(M), X_{\rm off,P}) P(M) dM,
    \label{eq:fxoff}
\end{equation}
where M$_{\rm low}$=1$\times$10$^{14}$M$_\odot$ and M$_{\rm up}$=8$\times$10$^{14}$M$_\odot$.  In addition, we marginalize over the detection probability as a function of mass $\hat{P}$(M). The mass trend is encoded in the root mean square variance of the density field $\sigma$(M). We calibrate such detection probability by dividing the eFEDS multiplicity function (see Eq. \ref{eq:Psigma}) computed from the cluster number density by the theoretical prediction and model it with an error function. We refer to the detection probability model as P(M), computed and modeled according to Eq. \ref{eq:Psigma}: 
\begin{align}
    f(\sigma) &= \frac{d\text{n}}{d\text{lnM}} \frac{M}{\rho_m}\Big(\frac{d \text{ln}\sigma^{-1}}{d \text{lnM}} \Big)^{-1}, \nonumber \\
    \hat{P}(M) &= f_{\rm eFEDS}(\sigma(M))/f_{\rm SEPPI+21}(\sigma(M)), \nonumber \\
    P(M) &= \frac{1}{2}erf[A(\log_{\rm 10}M_{\rm 500c} - M_{\rm 0})] + \frac{1}{2},
    \label{eq:Psigma}
\end{align}
where the parameters have been fit with the \textit{curve\_fit} software\footnote{\url{https://scipy.org}}. The values are M$_{\rm 0}$=14.314$\pm$0.001 and A=2.30$\pm$0.03. 
Finally, we account for projection effects by projecting the theoretical three dimensional $X_{\rm off,P}$ on the sky according to Eq. \ref{eq:Xoff_correction}:
\begin{equation}
    S_{X_{\rm off,P}} = X_{\rm off,P} \frac{1}{\pi} \int_0^\pi \sin{\theta}d\theta = \frac{2}{\pi} X_{\rm off,P}.
    \label{eq:Xoff_correction}
\end{equation}
We use the corrected S$_{X_{\rm off,P}}$ to compute the theoretical cumulative distribution function of the projected offset parameter.


\subsection{Prediction from hydrodynamical simulations}
\label{subsec:compare_with_hydro}
We process the Magneticum and TNG simulations in a similar way. For each halo in the simulation, we relate the optical center to the position of the main subhalo identified by the SubFind algorithm, which contains the central galaxy. We relate the X-ray center to the gas center computed from an emission measure weighted center of mass. The position of each gas particle contained by the halo is weighted by its mass and local density. We consider particles within the virial radius of each halo. We restrict to X-ray emitting gas particles with temperature between 0.1 and 10 keV. We finally compute the gas center according to Equation \ref{eq:emiss_weight_CM}:

\begin{align}
    w_{g,i} &= \rho_{g,i}\times m_{g,i}, \nonumber \\
    CM_{g} &= \frac{\sum_{\rm i=1}^{N} w_{g,i}\times x_{g,i}}{\sum_{\rm i=1}^{N} w_{g,i}},
    \label{eq:emiss_weight_CM}
\end{align}
where w$_{g,i}$ is the weight assigned to each gas particle, $\rho_{g,i}$ is the local gas density, m$_{g,i}$ is the gas particle mass, and x$_{g,i}$ is the particle position. The index i runs on the N gas particles contained by a halo.
We compute the offset between the CG and the gas center as their relative distance on the x--y projected cartesian plane.\\
In addition, we run the \textsc{rockstar} halo finder \citep[][]{Behroozi2013} on both Magneticum and TNG. We process the hydro simulation as well as the respective parent dark-matter-only runs. \textsc{rockstar} provides a measure of X$_{\rm off}$ for each identified halo. We focus on distinct main halos. We perform a positional matching between our \textsc{rockstar} halo catalogs from the hydro and DMO runs to the \textsc{subfind} catalogs provided together with the particle data from the Magneticum and TNG projects. The agreement between these catalogs is excellent. We discard halos where the location of the center disagrees by more than 500 kpc and the measure of M$_{\rm 500c}$ differs more than 10$\%$ between the three catalogs. Because the \textsc{subfind} catalogs provide the total mass, including stars and gas, we correct it by the gas fraction from \citet{Pratt2009A&A...498..361P}, see Fig. 8 therein, when comparing it to \textsc{rockstar} masses. The gas fraction from \citet{Pratt2009A&A...498..361P} provides a robust measure compared to a variety of other samples \citep[see][for a review]{Eckert2021AGN_feedback_review}. In total, we lose about 2$\%$ (1$\%$) of the halos with M$_{\rm 500c}$>1$\times$10$^{13}$ (1$\times$10$^{14}$) M$_\odot$. The matched halos allow us to compare the displacement between the gas center and the central galaxy to the offset parameter for common objects between the three catalogs. To match the average redshift of the high mass--low redshift eFEDS sample, we study the snapshot at z=0.30 for TNG-300. For Magneticum, we use the closest snapshot available at z=0.25, where the particle data has been stored for the full hydro and the DMO run. We verify that this does not bias our results in Sect. \ref{subsec:result_compare_to_sim}. The TNG-300 snapshot at z=0.30 contains 2232 (107) halos with M$_{\rm 500c}$>1$\times$10$^{13}$ (1$\times$10$^{14}$) M$_\odot$. The Magneticum Box2/hr snapshot at z=0.25 contains 9293 (314) halos with M$_{\rm 500c}$>1$\times$10$^{13}$ (1$\times$10$^{14}$) M$_\odot$.

\section{Results}
\label{sec:results}
In this Section, we present our main findings: the distribution of the displacement between the X-ray and optical center in eROSITA, its comparison to Magneticum and TNG, and to the N-body model from \citet{Seppi2021A&A...652A.155S}.

\subsection{Offset distributions and comparison to simulations}
\label{subsec:result_compare_to_sim}

\begin{figure*}[h]
    \centering
    \includegraphics[width=2.0\columnwidth]{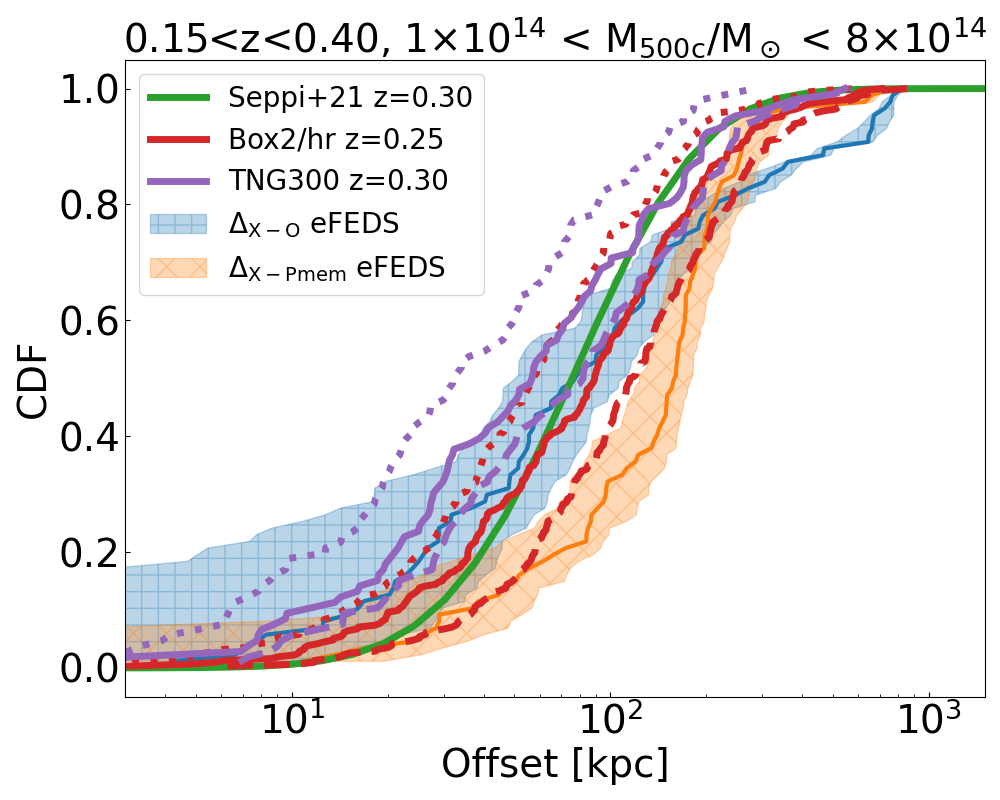}
	\caption{Comparison between the offsets measured in eROSITA, the prediction of the theoretical model, and hydrodynamical simulations. The cumulative distribution functions of the offsets between X-ray and optical centers for eFEDS clusters between redshift 0.15 and 0.4, and mass between 1$\times$10$^{14}$ and 8$\times$10$^{14}$ M$_\odot$ are denoted by the blue and orange lines. The first one refers to the optical center identified by the redMaPPer centering algorithm, the latter to the position of the galaxy with the largest membership probabilities. The shaded areas identify the uncertainty on the distributions. The green line shows the prediction obtained from the \citet{Seppi2021A&A...652A.155S} model described in Sect. \ref{subsec:seppi21_model}. The red (violet) one denotes the CDF of the offsets between the gas center and the CG position in the Magneticum (TNG) simulation described in Sect. \ref{subsec:compare_with_hydro}. The corresponding dashed and dotted lines account for the maximum and minimum projection effects. There is a broad agreement between the data, the prediction of the simulations, and the N-body model. However, the tails of the distributions are different. The N-body model predicts larger (smaller) displacements compared to data and hydrodynamical simulations at the low (high) offset end.}
    \label{fig:offset_withSIMS}
\end{figure*}

We focus first on the sample over which we have the most control: a subsample of 87 eFEDS clusters with 0.15 < z < 0.4 and 1$\times$10$^{14}$ < M$_{\rm 500c}$ < 8$\times$10$^{14}$ M$_\odot$. This is a secure sample with 340 counts per cluster on average. The redshift and the X-ray positional uncertainty have been measured for all these clusters. The average uncertainty on the X-ray position is 38 kpc. Given the mean values of M$_{\rm 500c}$=2.16$\times$10$^{14}$ M$_\odot$ and z=0.30 for this sample, the eFEDS selection function yields an average completeness of about 80$\%$ \citep[][]{2022A&A_LiuAng_eFEDS_clu}. The offsets $\Delta_{\rm X-O}$ and $\Delta_{\rm X-P_{\rm mem}}$ are shown by the blue and orange shaded areas in Fig. \ref{fig:offset_withSIMS}. The green line denotes the CDF of the projected offset parameter S$_{\rm X_{\rm off,P}}$, the red (violet) line shows the CDF of the offset in the Magneticum (TNG-300) simulation at z=0.25 (0.30). The corresponding dashed and dotted lines account for maximum and minimum projection effects. For each cluster in the simulations, we consider the largest possible displacement in the case where the two centers lay on a plane that is perpendicular to the line of sight, and the minimum one by choosing the smallest displacement after projecting the same clusters on the x-y, y-z, and x-z planes.
A comparison between observations and simulations at higher redshift requires deeper data, with accurate measurements of the X-ray position in the high-z regime.

\subsubsection{Average offsets}
For this specific sub-sample, we study the average offset at the 50$\%$ percentile point of the CDF. We measure $\Delta_{\rm X-O}$=76.3$_{\rm -27.1}^{\rm +30.1}$ kpc and $\Delta_{\rm X-P_{\rm mem}}$=157.4$_{\rm -34.0}^{\rm +20.6}$ kpc. The flattening of $\Delta_{\rm X-O}$ at large offsets is given by a tail of recent mergers and disturbed clusters. The average offset in hydrodynamical simulations is equal to 57.2 kpc for TNG-300 and 87.1 kpc for Magneticum Box2/hr. We see that both simulations predict offsets that are on average compatible with the distribution of $\Delta_{\rm X-O}$, but they are in disagreement with the offset between the X-ray center and the position of the galaxy with the largest membership probability $\Delta_{\rm X-P_{\rm mem}}$. Therefore, the displacement between the hot gas and the CG in hydrodynamical simulations is a good prediction of the offset between the optical center from redMaPPer and the X-ray position from eSASS in eROSITA data. To assess whether the different redshift considered for the Magneticum simulation impacts our findings, we do the same analysis for the snapshot at z=0.25 of the TNG-300 simulation, where the particle data is available also for the parent DMO run. We find that the offset is on average smaller by about 4 kpc compared to the snapshot at z=0.30. This is much smaller than the typical uncertainties on the data. We conclude that studying the snapshot at z=0.25 in the Magneticum simulation does not bias our results.  \\
The average value of the projected offset parameter is S$_{\rm X_{\rm off,P}}$=75.8 kpc. Similarly to the offsets predicted by TNG and Magneticum, it is in agreement with $\Delta_{\rm X-O}$, but disagrees with $\Delta_{\rm X-P_{\rm mem}}$. On average, we conclude that there is good agreement between the offsets measured in eROSITA clusters, the ones predicted by hydrodynamical simulations, and by the N-body model from \citet{Seppi2021A&A...652A.155S}. 

\subsubsection{Tails of the distributions}
The tails of the $\Delta_{\rm X-O}$ and S$_{\rm X_{\rm off,P}}$ distributions are different, where the CDF is smaller than about 0.2 and larger than 0.7. We attribute the discrepancy to baryonic effects such as dragging, cooling, and the disruption of the gas by AGN feedback and recent mergers. These effects tilt the shape of this distribution. This is in agreement with previous works, where the shape of the offset distribution changes from a modified Schechter function in N-body simulations \citep[][]{Rodriguez-Puebla2016, Seppi2021A&A...652A.155S} to a lognormal distribution in data \citep[][]{Mann2012MNRAS.420.2120M_offset}. We further discuss this result in Sect. \ref{sec:discussion}.

To separate relaxed and disturbed clusters, we follow the example of \citet{Ota2022_offset} and apply an offset cut according to
\begin{equation}
    \Delta_{\rm X-O} < 0.05 \times R_{\rm 500c}.
    \label{eq:relaxed_frac}
\end{equation}
We find that 27 clusters out of 87 are classified as relaxed. Our relaxed fraction of 31$\%$ is in agreement with the upper limit set at <39$\%$ by \citet{Ota2022_offset}. We additionally consider an upper limit of the relaxed fraction by accounting for the uncertainty on the measure of the offset (as explained in Sect. \ref{subsec:erosita_clusters}), assuming the lower limit of $\Delta_{\rm X-O}$ within the error. We apply again the same cut in Eq. \ref{eq:relaxed_frac} and obtain a relaxed fraction of < 59$\%$. Our results show that there is not a strong preference for relaxed objects in this eFEDS cluster sample. This is in agreement with previous work on eFEDS data. \citet{Ghirardini2022A&Amorph} combined eight different morphological parameters (central density, concentration, centroid shift, ellipticity, cuspiness, power ratios, photon asymmetry, and Gini coefficient) into the single relaxation score parameter. They did not find a clear preference for cool core clusters over disturbed ones and showed that the transition from a relaxed to a disturbed cluster population is smooth. In addition, \citet{Bulbul2022A&A_cluindisguise} analyzed the clusters hidden in the point-like sample of eFEDS sources, identifying them using optical data. They found that only the clusters in the point-like sample show a peaked profile in the central region. Finally, predictions from eROSITA simulations show that there is a significant preference for the detection of relaxed systems just in the point-like sample \citep[][]{Seppi2022A&A_eRASS1sim}.

\subsection{Full eROSITA samples}
\label{subsec:fullsamples}
We measure the position of the X-ray and optical centers for the 182 eFEDS and 4564 eRASS1 clusters as explained in Sect. \ref{sec:method}. We stress that our cuts in X-ray counts and richness discard faint clusters whose determination of the X-ray position is uncertain. An additional role in the location of the X-ray center is played by the cluster morphology. We use the relaxation score R$_{\rm score}$ measured by \citet{Ghirardini2022A&Amorph} on eFEDS clusters. They define clusters with R$_{\rm score}$>0.0019 as relaxed. Using the same R$_{\rm score}$ criterion, we measure an average uncertainty of 36 (64) kpc on the X-ray position for relaxed (unrelaxed) clusters. We conclude that the identification of the X-ray center is more secure for relaxed clusters.

\begin{figure}[h]
    \centering
    \includegraphics[width=1.0\columnwidth]{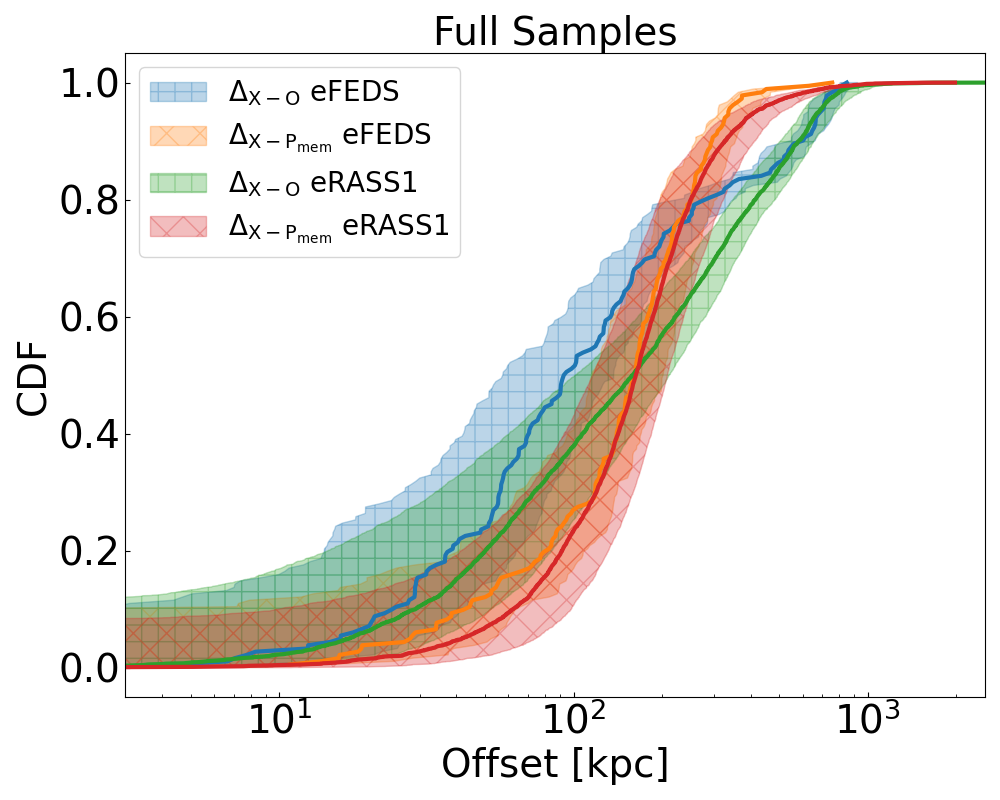}
	\caption{Cumulative distribution functions of the offsets between X-ray and optical centers for eROSITA clusters between redshift 0.15 and 0.8, more than 20 counts, and richness $\lambda$>20. These cuts yield 182 (4568) clusters from eFEDS (eRASS1). The shaded areas denote the uncertainty on the distributions. Different colors denote distinct definitions of the optical center: the one identified by the redMaPPer centering algorithm and the position of the galaxy with the largest membership probability (blue and orange for eFEDS, green and red for eRASS1).  }
    \label{fig:offset}
\end{figure}

We show the cumulative distribution function of the offsets in Figure \ref{fig:offset}. The blue (green) line shows the offset between the X-ray center and the redMaPPer center for the eFEDS (eRASS1) sample. The shaded areas denote the 1$\sigma$ error on the offset. We measure an average offset $\Delta_{\rm X-O}$=92.6$_{\rm -35.1}^{\rm +44.3}$ kpc in eFEDS and $\Delta_{\rm X-O}$=158.5$_{\rm -57.5}^{\rm +53.0}$ kpc in eRASS1. On average, the eRASS1 sample shows larger offsets compared to eFEDS. Since eRASS1 is a shallow survey compared to the deeper and more uniform eFEDS, and the detection probability for a given cluster grows as a function of exposure time \citep[][]{Clerc2018A&A...617A..92C, Seppi2022A&A_eRASS1sim}, it contains a larger fraction of high-mass, high-offset objects in the whole sample.  

\begin{figure}[h]
    \centering
    \includegraphics[width=1.0\columnwidth]{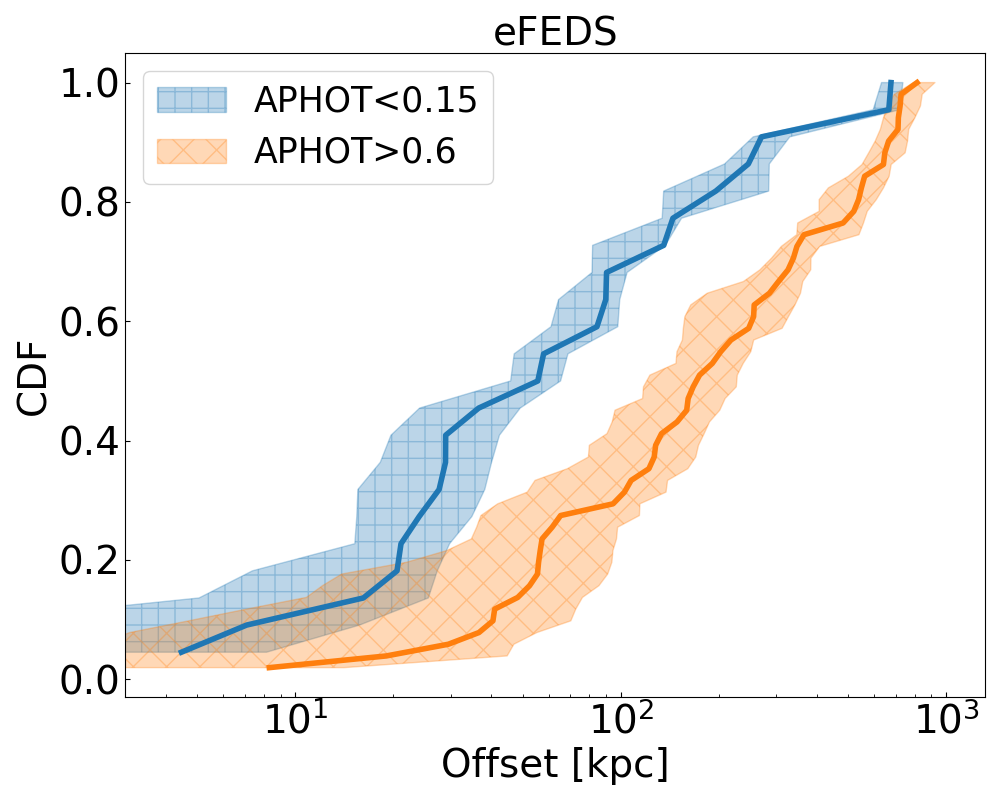}
	\caption{Cumulative distribution functions of the offsets between X-ray and optical centers for the eFEDS clusters. The blue shaded area denotes clusters classified as relaxed according to photon asymmetry APHOT<0.15, the orange one shows unrelaxed objects with APHOT>0.6.}
    \label{fig:offset_CC_NCC_eFEDS}
\end{figure}

The trends of the displacement between the X-ray center and the position of the galaxy with the largest membership probability are more similar between eFEDS and eRASS1 compared to $\Delta_{\rm X-O}$. They are shown by the orange (red) line in Fig. \ref{fig:offset} for eFEDS (eRASS1). The shaded areas denote the 1$\sigma$ error on the offset. We measure an average offset $\Delta_{\rm X-P_{\rm mem}}$=160.9$_{\rm -45.6}^{\rm +40.5}$ kpc in eFEDS and $\Delta_{\rm X-P_{\rm mem}}$=162.7$_{\rm -46.3}^{\rm +45.8}$ kpc in eRASS1.
The full samples mix clusters with different mass and redshift, which dilutes the intrinsic differences of $\Delta_{\rm X-P_{\rm mem}}$ between the eFEDS and the eRASS1 samples. \\
We find a weak correlation between the X-ray to optical offset $\Delta_{\rm X-O}$ and the cluster mass. We find a Pearson correlation coefficient equal to PCC=0.11 (0.08) for eFEDS (eRASS1). Similarly, we observe a weak correlation between the offset and redshift, with PCC=0.19 for eFEDS, and PCC=0.14 (0.10) for clusters with M$_{\rm 500c}$>1$\times$10$^{14}$ M$_\odot$ (groups with 1$\times$10$^{13}$<M$_{\rm 500c}$<1$\times$10$^{14}$ M$_\odot$) in eRASS1. The correlation is weak because of two contrasting effects. On the one hand, an isolated cluster relaxes in time, producing a small offset between X-ray and optical centers. On the other hand, mergers producing massive clusters create structures with complex morphology and large offsets. The positive correlation with increasing offset as a function of redshift is in agreement with the trend of X$_{\rm off}$ in N-body simulations \citep[][]{Seppi2021A&A...652A.155S}. The detection of more galaxy groups with low offset in future eROSITA all-sky surveys will shed light on the correlations between the offset and the cluster mass and redshift.\\
We study the offset for eFEDS clusters with different dynamical state, where morphological parameters were measured by \citet{Ghirardini2022A&Amorph}. \citet{Nurgaliev2013ApJAPHOT} showed that photon asymmetry is sensitive to the presence of substructure, an indication of morphological disturbance, also in a regime of low signal to noise ratio. This is useful for our study, given the relatively shallow eROSITA depth. In fact, we find that the photon asymmetry is the morphological parameter with the largest correlation to the X-ray to optical offset, with PCC=0.28. Following \citet[][]{Nurgaliev2013ApJAPHOT}, we separate relaxed clusters with APHOT<0.15 and unrelaxed ones with APHOT>0.6. We build the $\Delta_{\rm X-O}$ CDF for the two samples. The result is shown in Fig. \ref{fig:offset_CC_NCC_eFEDS}. We find an average offset of $\Delta_{\rm X-O}$=55.6$_{-9.7}^{+9.7}$ kpc for 22 clusters with APHOT<0.15. The offset for 51 clusters with APHOT>0.6 is larger $\Delta_{\rm X-O}$=174.4$_{-51.6}^{+53.3}$ kpc. We conclude that relaxed clusters exhibit a smaller offset compared to disturbed ones.

\section{Discussion}
\label{sec:discussion}
In this section, we discuss the offsets presented in Sect. \ref{sec:results}. The physical effects affecting the offsets play a key role in understanding the cause behind the smaller (larger) displacements measured in eROSITA data and hydrodynamical simulations compared to the N-body model in the low (high) offset regime.
The origin of the offsets in clusters of galaxies is related to the different response of each cluster component to different astrophysical phenomena. \\
The contribution of mergers and AGN feedback on the offset distribution is discussed in Sect. \ref{subsec:mergers} and \ref{subsec:AGN_feedback}. Their combination and the transition from the DMO scenario to the observed offset distribution are presented in Sect. \ref{subsec:offset_interpretation}. The discrepancies between the offsets predicted by different hydrodynamical simulations are discussed in Sect. \ref{subsec:offsets_in_sims}. Finally, the introduction of the offsets in a cosmological experiment is presented in Sect. \ref{subsec:cosmo_with_offsets}.

\subsection{Mergers}
\label{subsec:mergers}
Very large offsets likely originate from mergers between smaller objects into massive clusters. Mergers are one of the most energetic processes in the Universe, as the total kinetic energy involved reaches values up to 10$^{65}$ erg \citep{Markevitch1999ApJ...521..526M, Sarazin2002ASSL_merger, Markevitch2007PhR...443....1M}. In this context, it is particularly interesting to explore the differences between the central galaxy and the gas in relation to the dark matter distribution. 

The dark matter is mostly sensitive to gravitational interaction, while the CG and the gas are additionally subject to a variety of effects such as electromagnetic forces, ram pressure, scattering, and cooling \citep[][]{Merten2011MNRAS.417..333M}. When two clusters merge, the dark matter components stream through each other according to the evolving gravitational field, without being slowed down by the dragging experienced by baryonic components because of the additional interactions. The result is that after a merger, when the newly formed halo relaxes and the gas cools down, the offset between the dark matter profiles of the merging clusters is larger than the gas distribution one. In fact, clusters undergoing mergers typically show large offsets up to hundreds of kpc between different components \citep[][]{Menanteau2012ApJ_elGordo, Mann2012MNRAS.420.2120M_offset, Dawson2012ApJ...747L..42D, Monteiro-Oliveira2017MNRAS.466.2614M}. Large offsets provide therefore a hint of merger activity, compared to small offsets that characterize the pre-merger phase \citep{Jauzac2015MNRAS.446.4132J, Ogrean2015ApJ...812..153O}. An extreme case is the famous 1E 0657–56, also known as the bullet cluster \citep[][]{Markevitch2002ApJ_bullet, Clowe2006bullett}, where the total mass distribution traced by weak lensing extends to larger radii compared to the emission of the hot gas imaged with the Chandra X-ray observatory. This is in agreement with our result in Fig. \ref{fig:offset_withSIMS}, where we find a larger amount of clusters showing an offset of tens of kpc compared to the DMO prediction.

In addition, since the gas trails the dark matter during a merger because of ram pressure and friction, the gas starts sloshing within the cluster potential. This causes large offsets when the gas approaches the point of null velocity and positive acceleration during the sloshing process \citep[][]{Ascasibar2006ApJ...650..102A, Markevitch2007PhR...443....1M, Sanders2020A&A_pers_coma_slosh, Pasini2021ApJ...911...66P}. The complex behavior of the gas during the merging process is not easily mappable to the dark matter-only scenario. In these cases, the large offsets seen in data are not compatible with simple theoretical models, as described by \citet{DePropris2021MNRAS.500..310D}. The authors find that the BCG is generally aligned with the cluster mass distribution, showing that even if being displaced by a merger or if the dark matter halo is not relaxed, the central galaxy does follow the cluster potential. \citet{Hikage2018MNRAS.480.2689H} tested the performance of the redMaPPer centering algorithm using galaxy-galaxy lensing and confirmed that the central galaxy is not always the brightest member. A similar result was presented by \citet[][]{Hoshino2015MNRAS.452..998H}, who studied the distribution of luminous red galaxies in clusters. Therefore, the BCG is possibly a biased tracer of the deepest point of the halo potential well, especially for unrelaxed systems where the definition of the BCG is not trivial, and the brightest cluster galaxy may belong to a satellite merging halo. The identification of the central galaxy using centering probabilities with redMaPPer provides a better tracer of the center of the dark matter halo. This is in agreement with our results. In fact, the median of the $\Delta_{\rm X-P_{\rm mem}}$ distribution does not agree with the DMO model, with hydrodynamical simulations, nor with the median of $\Delta_{\rm X-O}$. The additional information from the whole galaxy population encoded in $\Delta_{\rm X-O}$ provides an optical center that is on average closer to the X-ray center. However, in complex mergers, the contribution of galaxies extending to the cluster outskirts may shift the optical center away from the X-ray one compared to the galaxy with largest membership probability alone. This explains the extension to large offsets for $\Delta_{\rm X-O}$ in the most disturbed clusters. \\
Compared to the definition of the optical center, the X-ray emitting gas may not properly trace the center of the cluster potential after being disrupted by complex mergers. This was studied by \citet[][]{Cui2016bcg}. The authors analyzed the location of different centers of galaxy clusters in simulations and found that the BCG shows a better correlation to the center of the gravitational potential compared to the X-ray gas. They measure an average offset between the BCG and the potential center smaller than 10 kpc. The displacement between X-ray and potential centers reaches average values of tens of kpc. This is also in agreement with Fig. \ref{fig:offset_withSIMS}, where the data shows larger offsets compared to the DMO prediction in the high offset regime. We conclude that on one hand, the central galaxy is on average more likely to be trapped in the vicinity of the deepest point of the cluster potential. On the other hand, the X-ray center, being related to gas permeating the whole halo, is more sensitive to the overall variations of the potential during merger activity and is altered by AGN feedback (see Sect. \ref{subsec:AGN_feedback}). This is in concordance with previous work on observations \citep[][]{George2012centering} and simulations \citep[][]{Cui2016bcg}. 

\subsection{AGN feedback}
\label{subsec:AGN_feedback}
The AGN feedback plays an additional role in this context. Efficient accretion onto the SMBH of the central galaxy is known to impact the gas on very large scales inside the dark matter halo hosting a galaxy cluster \citep[see][for a review]{Eckert2021AGN_feedback_review}. The central AGN does not only reorganize the gas on large scales but the presence of jets digging cavities in the gas distribution produces a significant diversity of the gas morphology. The structure of the gas can be disrupted by AGN feedback, which pushes the gas away from the cluster center \citep[][]{Gaspari2012ApJ...746...94G, Gitti2012_CCreview, McNamara2012NJPh_feedback, Li2015ApJ_feedback}. This contributes to the larger offsets measured in the presence of baryons compared to the N-body simulations. However, the AGN impact on the offsets is not immediate. In fact, the central galaxy becomes active when there is enough gas supply to the central region of the cluster, which means that a cluster is more likely to be relaxed shortly before the beginning of AGN activity \citep[][]{Fabian2012ARA&A_feedback, Pinto2018MNRAS_phoenix_feedback}. It is also reasonable to expect a correlation between the AGN impact on the gas distribution and redshift. Dark matter halos are smaller at early times and the feedback may distribute and reorganize the gas on large scales more easily. \\
The notion that the gas is displaced compared to the dark matter distribution has been explored by \citet{Cui2016bcg}. They compare simulations with and without AGN feedback and find that its activation enhances the offset between the gas and the dark matter centers, especially for clusters with offsets between around 10 and 30 kpc in the simulation without AGN. The offset reaches an average value of about 70 kpc in the run with active AGN. The authors also demonstrate that the X-ray centroid is more consistent than the X-ray peak between hydrodynamical runs with different baryonic physics. This is also supporting our way of locating the X-ray center with eSASS, which accounts for the overall distribution of the emission, rather than simply choosing the brightest pixel.

\subsection{Physical interpretation of the offset distribution}
\label{subsec:offset_interpretation}

\begin{figure}[h]
    \centering
    \includegraphics[width=1.0\columnwidth]{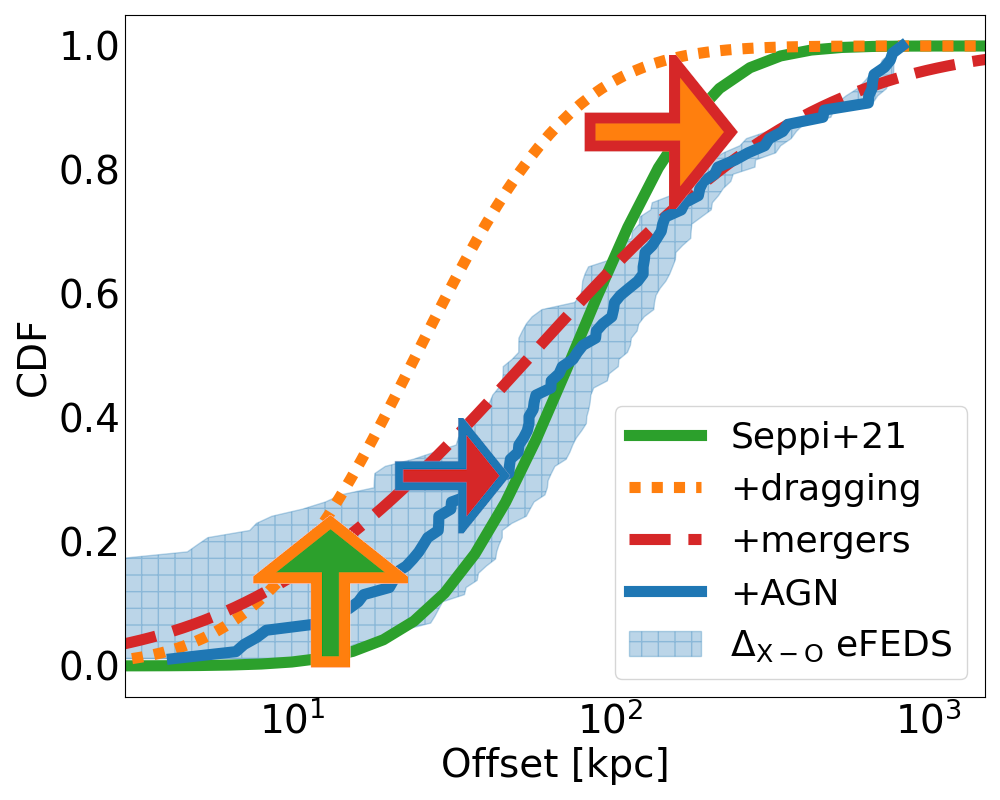}
	\caption{Illustration showing the interpretation of the impact of different astrophysical effects on the offset distribution. The green line refers to the dark matter only scenario (see Fig. \ref{fig:offset_withSIMS}). The dragging due to ram pressure and baryon friction increases the number of clusters with a small offset and is displayed in orange. This first transition is highlighted by the green arrow with an orange edge. Major and minor mergers are responsible for the largest offsets, which further shift the right-hand tail of the distribution (in red). This second transition is highlighted by the orange line with a red edge. Finally, AGN feedback increases the offsets in a medium regime, reducing the number of clusters with a small offset. The final result is the $\Delta_{\rm X-O}$ distribution measured in eFEDS (see Fig. \ref{fig:offset_withSIMS}). It is shown in blue and the third transition is displayed by the red arrow with a blue edge.}
    \label{fig:offset_cartoon}
\end{figure}

We combine the discussion from the previous paragraphs and formulate a physically motivated interpretation of the offset distributions in Fig. \ref{fig:offset_withSIMS}. We use the illustration in Fig. \ref{fig:offset_cartoon} to qualitatively guide the discussion. The green line showing the DMO analytical model, and the blue line with shaded area denoting the eFEDS result are the same as in Fig. \ref{fig:offset_withSIMS}. We interpret the shift of the distribution from the DMO scenario to the observations due to different astrophysical phenomena. First, the addition of small scale baryonic effects such as dragging, ram pressure, and friction reduces the offsets compared to the DMO case. This is likely to happen in minor mergers, where the gas distribution is not catastrophically disrupted, but the gas ends up trailing the dark matter component of the merging objects. The baryon dragging is reflected in an increment of the CDF at small offsets (orange line), shown by the green arrow with an orange edge. In addition, complex and major mergers can significantly disrupt the gas distribution or even strip the central galaxy from the bottom of the potential well, resulting in larger offsets compared to the DMO scenario. This causes a shift of the right-hand side of the CDF towards larger offsets, from the green and orange lines to the red one. The transition is highlighted by the orange arrow with a red edge. Furthermore, the AGN feedback alters the gas distribution, reducing the number of clusters with a small offset, as shown by \citet[][]{Cui2016bcg}. The final CDF is therefore more skewed towards larger offsets, following the red arrow with a blue edge. The final result is the offset distribution measured in the eFEDS subsample. It includes all these contributions and is shown by the blue line. The final CDF grows less rapidly compared to the dark matter only case, which is what we find when comparing $\Delta_{\rm X-O}$ to the analytical DMO model (see Fig. \ref{fig:offset_withSIMS}). Very large samples in future eROSITA all-sky surveys will allow a more detailed study of cluster relaxation and offsets at fixed cluster properties such as mass and redshift.

\subsection{Discrepancy between offsets in simulations}
\label{subsec:offsets_in_sims}
The different offsets predicted by TNG and Magneticum may be caused by different aspects. First, the Magneticum and TNG simulations are run assuming different cosmologies (see Table \ref{tab:simulation_parameters}). In particular, the WMAP cosmology assumed for Magneticum is slower in producing collapsed structures, due to the smaller $\Omega_{\rm M}$ and $\sigma_{\rm 8}$ compared to the Planck cosmology in TNG. Therefore, at a fixed redshift, the merger rate is different between the two simulations: halos have merged more recently in Magneticum because the growth factor is proportional to the matter density in the Universe. It makes them more disturbed, which may additionally contribute to the larger offsets predicted by Magneticum compared to TNG.  

An additional factor is the AGN feedback scheme \citep[][]{Hirschmann2014MNRAS.442.2304H, Weinberger2018MNRAS_TNG_AGNfeedback}. The basic structure of the accretion onto SMBHs is similar in these two simulations. It is based on an Eddington-limited Bondi accretion rate, following the Bondi–Hoyle–Lyttleton approximation \citep[][]{Bondi1944MNRAS.104..273B, Bondi1952MNRAS.112..195B}, and accounts for a two-way accretion mode, transitioning from a high accretion state (quasar mode), characterized by the presence of a thin disk, where the feedback is inefficient and released into the surrounding gas as thermal energy, to a low accretion state (radio mode), characterized by the quiescent infall of gas from the hot halo in quasi-hydrostatic equilibrium. In this case, the feedback is more efficient, and powerful radio jets are produced, that heat the gas kinetically \citep[see][]{Croton2006MNRAS.365...11C, Fanidakis2011MNRAS.410...53F}. \\
AGN feedback models reproduce the majority of AGN observations but struggle to perfectly grasp the full wealth of observed properties \citep[][]{Biffi2018lightcone, Comparat2019agn}. Detailed predictions should therefore be taken with caution. Nonetheless, different choices of the parameters in the feedback prescription may explain the larger offsets predicted by Magneticum compared to TNG. For example, the transition between the quasar mode and the radio mode, based on a choice of the Eddington ratio between accretion rate and Eddington limit, follows different thresholds. This is fixed at 1$\%$ in Magneticum. In TNG instead, a black hole mass-dependent threshold is chosen, such that its value is smaller than 1$\%$ for M$_{\rm BH} \lessapprox$10$^{8.4}$ M$_\odot$, and can reach larger values of 10$\%$ only for the most massive black holes of 10$^9$ M$_\odot$. Therefore, the radio mode where the feedback is more efficient is active for longer accretion phases in Magneticum compared to TNG. The gas may be ultimately pushed out to smaller distances in TNG, causing the lower values of the offsets. In addition, other differences may impact the modeling of AGN feedback in relation to the offsets. For example, the feedback efficiency in the thermal mode is slightly larger in Magneticum (0.03) than TNG (0.02). The efficiency in the kinetic mode is fixed in Magneticum (0.1), while in TNG it depends on the local density of the environment, which makes the coupling between AGN feedback and gas weaker in low density regions. In both cases, the gas may experience a push out to larger distances in Magneticum. These different prescriptions lead to a redshift-dependent switch of the feedback mode in TNG. \\
Moreover, a black hole seed of mass 1.18$\times$10$^{6}$ M$_\odot$ is generated at the center of dark matter halos more massive than 7.38$\times$10$^{10}$ M$_\odot$ in TNG \citep[][]{Weinberger2017MNRAS.465.3291W}. In Magneticum instead, the assignment is based on the stellar mass of the halo, as a seed black hole of 4.55$\times$10$^{6}$ M$_\odot$ is placed at the position of the most bound stellar particle in halos with stellar mass $M_\star \gtrapprox$1.4$\times$10$^{10}$ M$_\odot$. Finally, Magneticum also allows the accretion of fractions of each gas particle onto the SMBH, providing a more continuous representation of the accretion process. \\
The X-ray to optical offset is ultimately dependent on the efficiency and the geometry of AGN feedback, rather than individual parameters of the specific feedback implementation to individual simulations. This makes our results robust on various feedback receipts in different simulations. The X-ray to optical offset may be used as a diagnostic quantity to inform AGN feedback models in the future generation of large simulations with baryons. \\
Our findings are in agreement with independent work on the suppression of the matter power spectrum due to baryons \citep{Chisari2018MNRAS_baryons, Arico2021MNRAS_baryons}. In particular, \citet{Schneider2019JCAP_baryons} calibrated a model of baryonic effects on the power spectrum using the X-ray baryon fraction. They show that their calibration of the power spectrum suppression on small scales (k $>\sim$ 1 h/Mpc) is stronger compared to the prediction by the TNG simulations. This is in agreement with our result of smaller average offsets in TNG compared to Magneticum. A systematic study of a larger amount of simulations is needed to quantitatively assess this effect. An example is provided by the CAMELS project \citep[][]{Villaescusa2021ApJCAMEL}. They produce a suite of 4233 simulations with different cosmologies and varying stellar and AGN feedback prescriptions. However, the limited size of such boxes (25 Mpc/h) does not allow a quantitative study for clusters of galaxies.

\subsection{Cosmology with offsets}
\label{subsec:cosmo_with_offsets}

\begin{figure}[h]
    \centering
    \includegraphics[width=1.0\columnwidth]{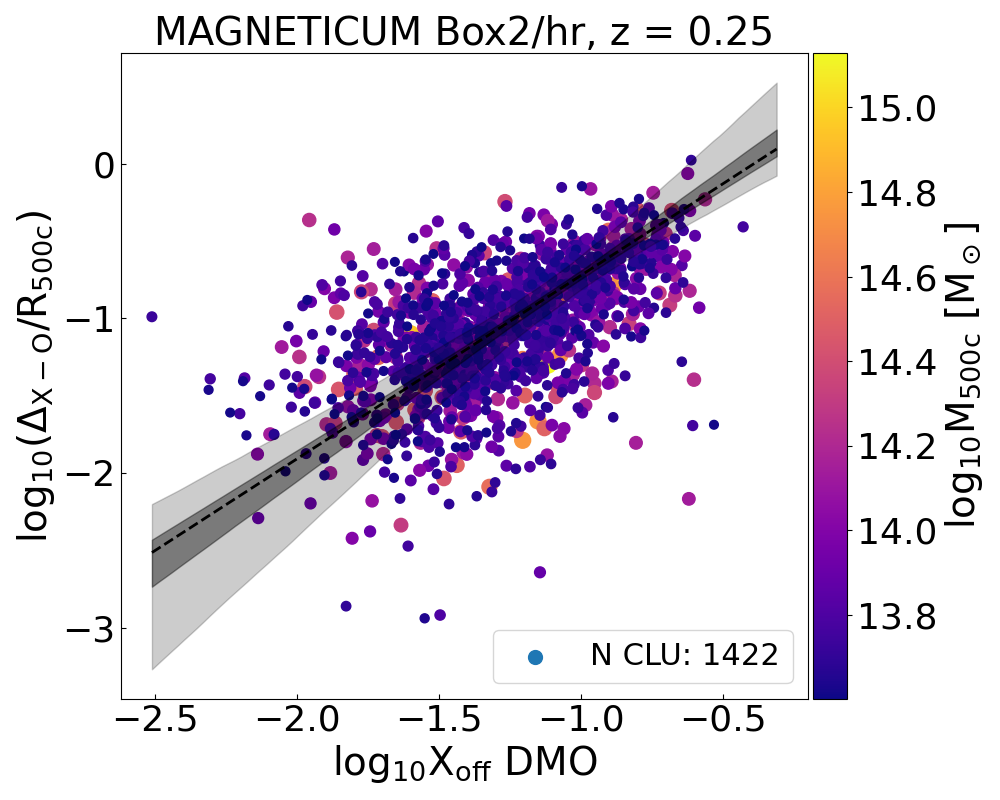}
    \includegraphics[width=1.0\columnwidth]{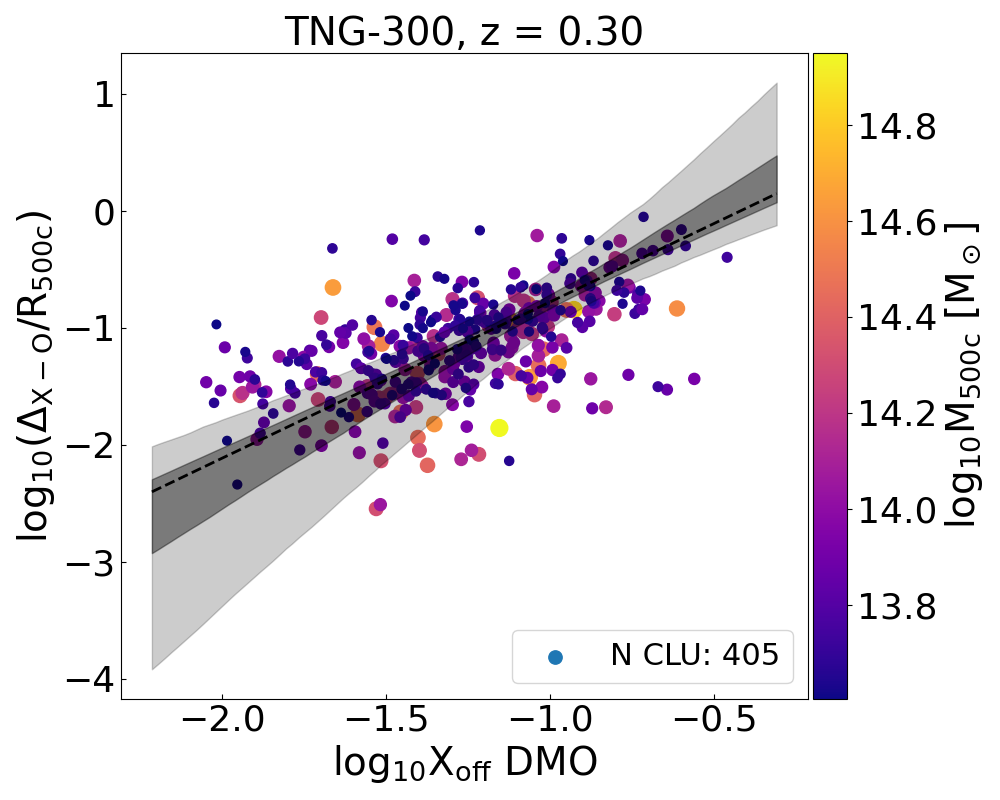}    
	\caption{Relation between the displacement $\Delta_{\rm X-O}$ and the offset parameter X$_{\rm off}$. The upper panel shows the Magneticum Box2/hr simulation, the bottom panel refers to TNG-300. Each dot denotes a halo. The dots are color-coded as a function of mass. The black dashed line shows the best-fit model, the black shaded areas denote the 1$\sigma$ and 2$\sigma$ uncertainty on the model.}
    \label{fig:offset_xoff_relation}
\end{figure}

\begin{figure}
    \centering
    \includegraphics[width=1.0\columnwidth]{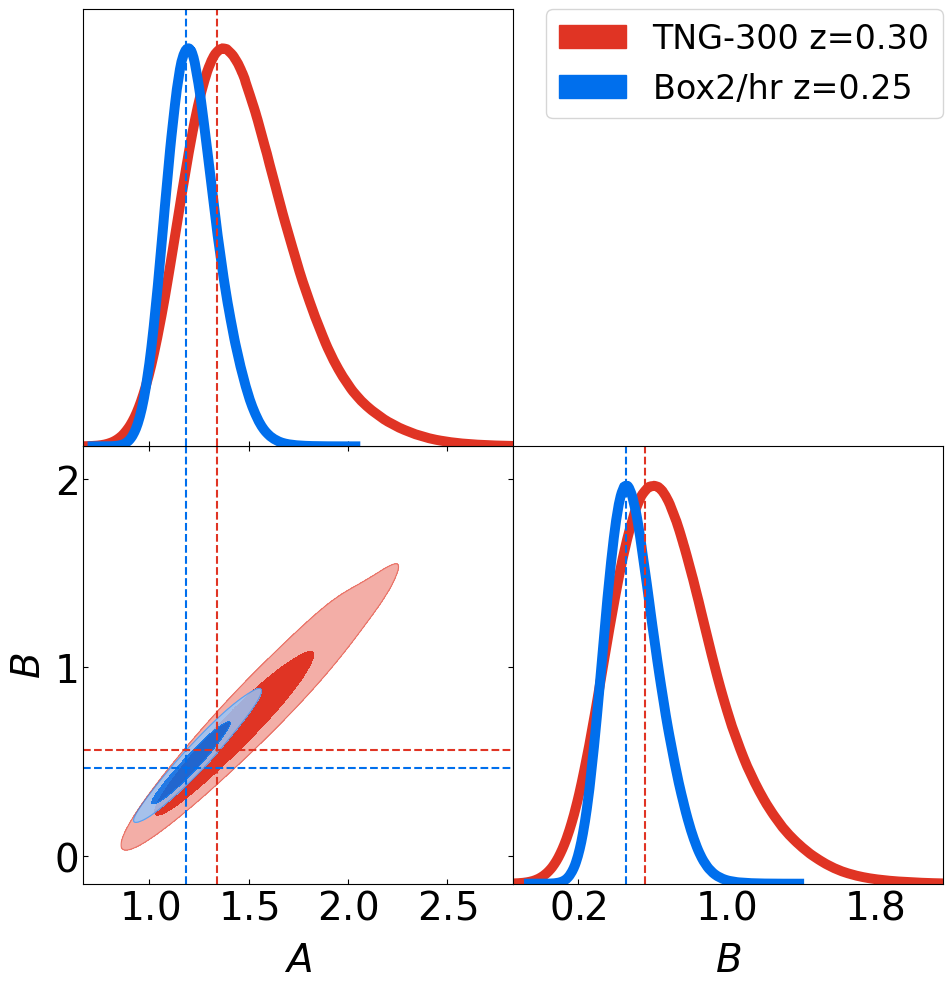}
    \caption{Triangular plot showing the posterior distributions of the best-fit parameters relating $X_{\rm off}$ to the offset between the gas center and the central galaxy in simulations (see Eq. \ref{eq:offset_xoff}). The red (blue) lines and contours show the TNG-300 (Magneticum-Box2) simulation. The shaded areas of the bottom-left panel denote the 1$\sigma$ and 2$\sigma$ confidence level contours.}
    \label{fig:bestfit_contours}
\end{figure}

The halo mass function model developed by \citet{Seppi2021A&A...652A.155S} allows marginalizing the halo abundance on variables related to the dynamical state of dark matter halos, such as $X_{\rm off}$. This mitigates related selection effects. For example, some X-ray-selected cluster samples are affected by the cool core bias \citep[][]{Eckert2011}. Relaxed clusters where the gas has cooled in the central region exhibit a peaked emission in the core. It potentially biases the detection towards such objects, compared to non-cool core ones, where the emissivity profile is flatter. We propose to use the offset between X-ray and optical centers as an observable to link real data to $X_{\rm off}$. This has the potential to enable a cosmological cluster count experiment as a function of mass and offset, unbiased by selection effects related to the cluster dynamical state. \\
We study the relation between $\Delta_{\rm X-O}$/R$_{\rm 500c}$ in simulations and the offset parameter X$_{\rm off}$. We use the value of X$_{\rm off}$ measured with \textsc{rockstar} on the halos in the DMO parent simulation. This connects observable properties related to the gas and stars in clusters to intrinsic properties of halos in the model calibrated on N-body simulations \citep[][]{Seppi2021A&A...652A.155S}. There is a positive correlation between $\Delta_{\rm X-O}$ and X$_{\rm off}$. Disturbed clusters with a large offset parameter in N-body simulations also exhibit a large displacement between the gas and the central galaxy in the respective hydro run. 
We model the correlation between the offset measured in Magneticum and TNG to X$_{\rm off}$ with a power-law relation (Eq. \ref{eq:offset_xoff}):

\begin{equation}
    \log_{10}\dfrac{\Delta_{\rm X-O}}{R_{\rm 500c}} = A\times \log_{10}X_{\rm off} + B.
    \label{eq:offset_xoff}
\end{equation}

\begin{table}[]
    \centering
    \caption{Best fit parameters for the relation between the displacement between the gas and the central galaxy and the offset parameter in simulations.}
    \begin{tabular}{|c|c|c|}
        \hline
        \rule{0pt}{2.2ex} & A & B  \\
        \hline
        \rule{0pt}{2.2ex} Magneticum Box2/hr & 1.19$\pm$0.24 & 0.46$\pm$0.27 \\
        TNG-300-1 & 1.34$\pm$0.59 & 0.56$\pm$0.63 \\
        \hline
    \end{tabular}
    \label{tab:offset_xoff_bestfit}
\end{table}

In Sect. \ref{subsec:fullsamples} we found a weak correlation between cluster masses and offset. However, there is no specific trend as a function of mass in Eq. \ref{eq:offset_xoff}. Indeed the mass dependence is in the normalization of $\Delta_{\rm X-O}$ to R$_{\rm 500c}$ and of the offset between the center of mass and the peak of the halo profile to the virial radius. Therefore, we fit halos of different masses together. We perform the fitting using the UltraNest\footnote{\url{https://johannesbuchner.github.io/UltraNest/}} package \citep[][]{Buchner2019, Buchner2021_ultranest}. We fit all individual halos more massive than M$_{\rm 500c}$>4$\times$10$^{13}$ M$_\odot$. We assume a Poisson likelihood of the form $\log{\mathcal{L}} = -\sum M + \sum D \times \log M$, 
where $M$, $D$ represent the model and the data, respectively. The $\Delta_{\rm X-O}$/R$_{\rm 500c}$ to X$_{\rm off}$ relation is presented in Fig. \ref{fig:offset_xoff_relation}. The top (bottom) panel shows the Magneticum (TNG) simulation. The figure is color-coded according to mass, spanning from low mass groups with M$_{\rm 500c}$=4$\times$10$^{13}$ M$_\odot$ to the most massive clusters with M$_{\rm 500c}$>1$\times$10$^{15}$ M$_\odot$. The black dashed lines show the best-fit model and the black shaded areas contain the 1$\sigma$ and 2$\sigma$ uncertainties on the model. The Magneticum simulation shows a tighter constraint on the relation compared to TNG. Indeed, Magneticum contains a larger amount of halos than TNG thanks to its larger volume, which enables more precise modeling of the $\Delta_{\rm X-O}$--$X_{\rm off}$ relation. The best fit parameters are shown in Table \ref{tab:offset_xoff_bestfit}. The slope and the normalization of the relation (parameters $A$ and $B$) are compatible between Magneticum and TNG. The full 2D and the marginalized 1D posterior distributions are shown in Fig. \ref{fig:bestfit_contours}. The blue contours denoting the Magneticum simulation span a smaller area on the A-B plane compared to the red ones, showing the TNG-300 box, because of the different amount of halos in the two simulations. 

\begin{figure}
    \centering
    \includegraphics[width=1.0\columnwidth]{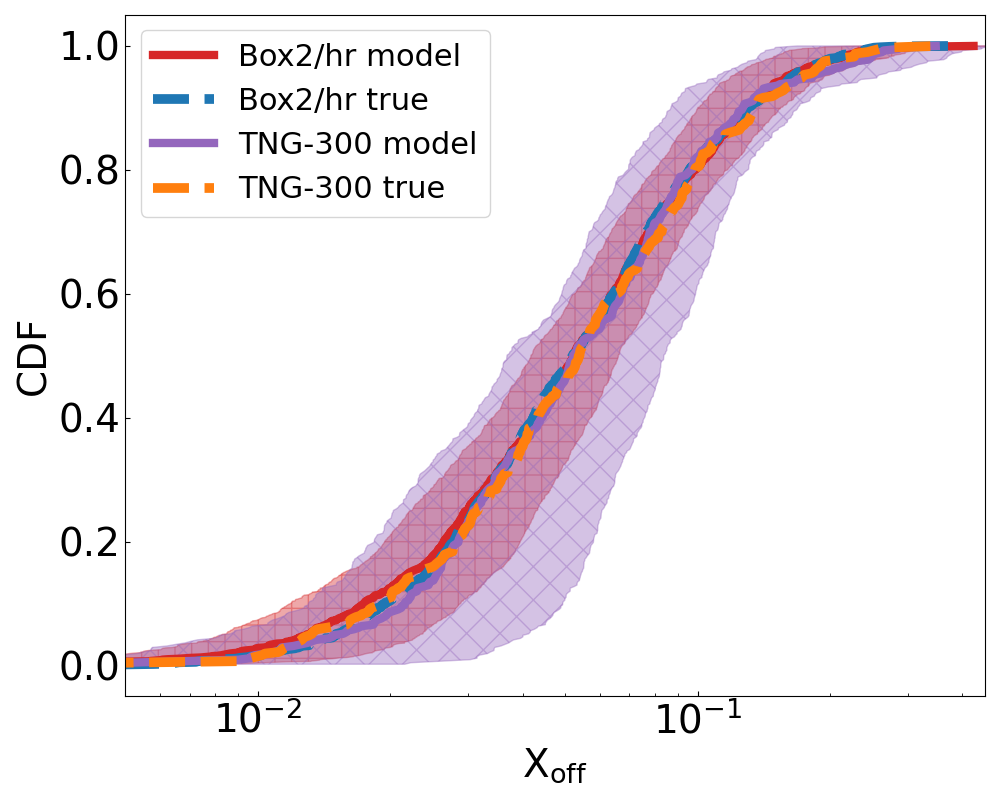}
    \caption{Recovery of the theoretical $X_{\rm off}$ distribution using the $\Delta_{\rm X-O}$--$X_{\rm off}$ relation. The application of Eq. \ref{eq:offset_xoff} on the offset measured in Magneticum and TNG is shown in red and violet, respectively. The shaded areas contain the 1$\sigma$ uncertainty on the model of the $\Delta_{\rm X-O}$--$X_{\rm off}$ relation. The blue (orange) dashed line refer to the direct measure of X$_{\rm off}$ in the DMO counterpart of Magneticum (TNG). We find excellent agreement between the distribution of the true X$_{\rm off}$ and the prediction of Eq. \ref{eq:offset_xoff}.}
    \label{fig:recovery_of_xoff}
\end{figure}

Starting from the measure of $\Delta_{\rm X-O}$/R$_{\rm 500c}$ in the simulations, we predict the X$_{\rm off}$ distribution in Magneticum and TNG by inverting the model in Eq. \ref{eq:offset_xoff}. We find an excellent agreement between the distribution of X$_{\rm off}$ measured on the DMO counterparts of the hydrodynamical simulations with the prediction obtained from the X-ray to optical offset and inverting Eq. \ref{eq:offset_xoff}. The result is shown in Fig. \ref{fig:recovery_of_xoff}. The true X$_{\rm off}$ CDF is shown in blue for Magneticum and in orange for TNG. The red and violet lines denote the prediction of X$_{\rm off}$ from $\Delta_{\rm X-O}$. The shaded areas include the 1$\sigma$ uncertainty on the model in Eq. \ref{eq:offset_xoff}. Our prediction of X$_{\rm off}$ from the X-ray to optical offset is able to recover the true X$_{\rm off}$ distribution with great precision. It enables the direct mapping of an observable offset to the offset parameter in DMO simulations, providing a reliable estimator of the mass--X$_{\rm off}$ function g($\sigma(M)$, X$_{\rm off}$) \citep[][]{Seppi2021A&A...652A.155S}.\\
In a full end-to-end cosmological study as a function of cluster mass and offset, one can marginalize over the parameters of the relation in Eq. \ref{eq:offset_xoff}, similarly to the standard way of marginalizing over the mass observable scaling relation parameters in recent cosmological analysis with clusters of galaxies \citep[][]{Mantz2015cosmology, Bocquet2019spt_cosmo, IderChitham2020MNRAS.499.4768I}. Finally, the similar correlation of $\Delta_{\rm X-O}$ and X$_{\rm off}$ with redshift explained in Sect. \ref{subsec:fullsamples} is key for future studies modeling the redshift trend of the $\Delta_{\rm X-O}$--X$_{\rm off}$ relation.

\section{Summary and Conclusion}
\label{sec:conclusions}
The eROSITA X-ray telescope is detecting clusters of galaxies at an unprecedented rate. It provides a large sample of clusters and groups to study astrophysical properties and constrain cosmological parameters. A key aspect of galaxy cluster studies is the definition of their center.

In this work, we measure the offset between the X-ray and optical centers for clusters observed by eROSITA. We consider two cluster catalogs: the eFEDS and the eRASS1 samples. We study two possible definitions of the optical center: the one provided by the centering algorithm of redMaPPer, and the position of the member galaxy with the largest membership probability. On average, the offsets measured in eRASS1 are larger compared to eFEDS (see Fig. \ref{fig:offset}). This is a consequence of the shallower exposure of eRASS1. It does not allow detecting lower mass clusters with smaller offsets, that are instead present in the eFEDS sample. For eFEDS, we use the morphological measured by \citet[][]{Ghirardini2022A&Amorph} and find that the offset correlates best with photon asymmetry. However, a quantitative interpretation of the offsets distributions for the whole samples is complicated, because they include clusters at different evolutionary phases, with various mass and redshift. \\
Therefore, we select a well controlled subsample of eFEDS (0.15 < z < 0.4 and 1$\times$10$^{14}$ < M$_{\rm 500c}$ < 8$\times$10$^{14}$ M$_\odot$), where the masses have been measured using weak lensing. We measure an average offset $\Delta_{\rm X-O}$=76.3$_{\rm -27.1}^{\rm +30.1}$ kpc. Using a threshold of $\Delta_{\rm X-O}$ < 0.05 $\times$ R$_{\rm 500c}$ \citep[][]{Ota2022_offset} to select relaxed systems, we measure a relaxed fraction of 31$\%$. According to this criterion, the eFEDS subsample does not show a preference for relaxed clusters, in agreement with \citet{Ghirardini2022A&Amorph, Bulbul2022A&A_cluindisguise}.\\
We compare the offsets measured in such sample to the ones predicted by the Magneticum Box2/hr and TNG-300 hydrodynamical simulations, and the N-body model of the offset parameter from \citet{Seppi2021A&A...652A.155S}. In the hydrodynamical simulations, we locate the optical center at the position of the main subhalo and the X-ray center at the center of mass of the hot gas particles weighted by their mass and density. The result is shown in Fig. \ref{fig:offset_withSIMS}. We find a broad agreement between them, especially for the median of the distribution (i.e., the 50$\%$ percentile point of the CDF). However, the tails of the distributions are different.
We find that the offsets measured in eROSITA data and predicted by Magneticum and TNG are smaller (larger) compared to the N-body model in the low (high) offset regime. This inconsistency is caused by baryons, that reduce the offsets for relaxed systems due to cooling and dragging and increase it for disturbed ones, mainly due to mergers and secondly AGN feedback. This scenario agrees with other work on observations \citep[][]{Churazov2003ApJ_Perseus, George2012centering, Zenteno2020MNRAS.495..705Z} and simulations \citep[][]{Molnar2012ApJ...748...45M, Zhang2014ApJ_mergerSIM, Cui2016bcg}. 
We also find that considering the optical center provided by redMaPPer instead of the galaxy with the largest membership probability provides a better comparison with simulations.
Deeper eROSITA all-sky surveys will allow studying the offsets for a population of clusters at high redshift, which is currently limited by the relatively small area of eFEDS and the shallow depth of eRASS1. Larger eRASS cluster samples will also allow investigating cross-correlations of the offset with mass and redshift.

We explore the possibility of introducing the offsets in a cosmological cluster count experiment. We use the displacement between X-ray and optical centers as a proxy for $X_{\rm off}$. We fit a power-law relation between them (see Figs. \ref{fig:offset_xoff_relation} and \ref{fig:bestfit_contours}). We find a relation that is independent of mass, thanks to the normalization to the cluster size ($R_{\rm 500c}$ for $\Delta_{\rm X-O}$ and R$_{\rm vir}$ for X$_{\rm off}$). Our model of the $\Delta_{\rm X-O}$--X$_{\rm off}$ relation provides a precise prediction of the true X$_{\rm off}$ distribution in Mangeticum and TNG. It is then possible to measure the cluster abundance as a function of mass and offset and use the model from \citet{Seppi2021A&A...652A.155S} to constrain cosmological parameters. The best-fit parameters of the $\Delta_{\rm X-O}$--$X_{\rm off}$ relation can be marginalized over similarly to the common mass observable scaling relation. This allows marginalizing over selection effects related to the cluster dynamical state directly in the measure of the halo mass function.

\section*{Acknowledgements}
This work is based on data from eROSITA, the soft X-ray instrument aboard SRG, a joint Russian-German science mission supported by the Russian Space Agency (Roskosmos), in the interests of the Russian Academy of Sciences represented by its Space Research Institute (IKI), and the Deutsches Zentrum für Luft- und Raumfahrt (DLR). The SRG spacecraft was built by Lavochkin Association (NPOL) and its subcontractors, and is operated by NPOL with support from the Max Planck Institute for Extraterrestrial Physics (MPE).

The development and construction of the eROSITA X-ray instrument was led by MPE, with contributions from the Dr. Karl Remeis Observatory Bamberg \& ECAP (FAU Erlangen-Nuernberg), the University of Hamburg Observatory, the Leibniz Institute for Astrophysics Potsdam (AIP), and the Institute for Astronomy and Astrophysics of the University of Tübingen, with the support of DLR and the Max Planck Society. The Argelander Institute for Astronomy of the University of Bonn and the Ludwig Maximilians Universität Munich also participated in the science preparation for eROSITA. 

The eROSITA data shown here were processed using the eSASS/NRTA software system developed by the German eROSITA consortium.

The authors gratefully acknowledge the Gauss Centre for Supercomputing e.V. (\url{www.gauss-centre.eu}) for funding this project by providing computing time on the GCS Supercomputer SuperMUC at Leibniz Supercomputing Centre (\url{www.lrz.de}).

VB acknowledges support by the \emph{Deutsche Forschungsgemeinschaft, DFG} project nr. 415510302. KD acknowledges support by the DFG under Germany’s Excellence Strategy - EXC-2094 - 390783311 as well as through the COMPLEX project from the European Research Council (ERC) under the European Union’s Horizon 2020 research and innovation program grant agreement ERC-2019-AdG 882679. The calculations for the {\it Magneticum} hydrodynamical simulations were carried out at the Leibniz Supercomputer Center (LRZ) under the project pr83li. We are especially grateful for the support by M. Petkova through the Computational Center for Particle and Astrophysics (C2PAP).

The authors thank the anonymous referee for the constructive comments about this manuscript.

\appendix

\section{Images}
\begin{figure*}[ht]
    \centering
    \includegraphics[width=0.73\columnwidth]{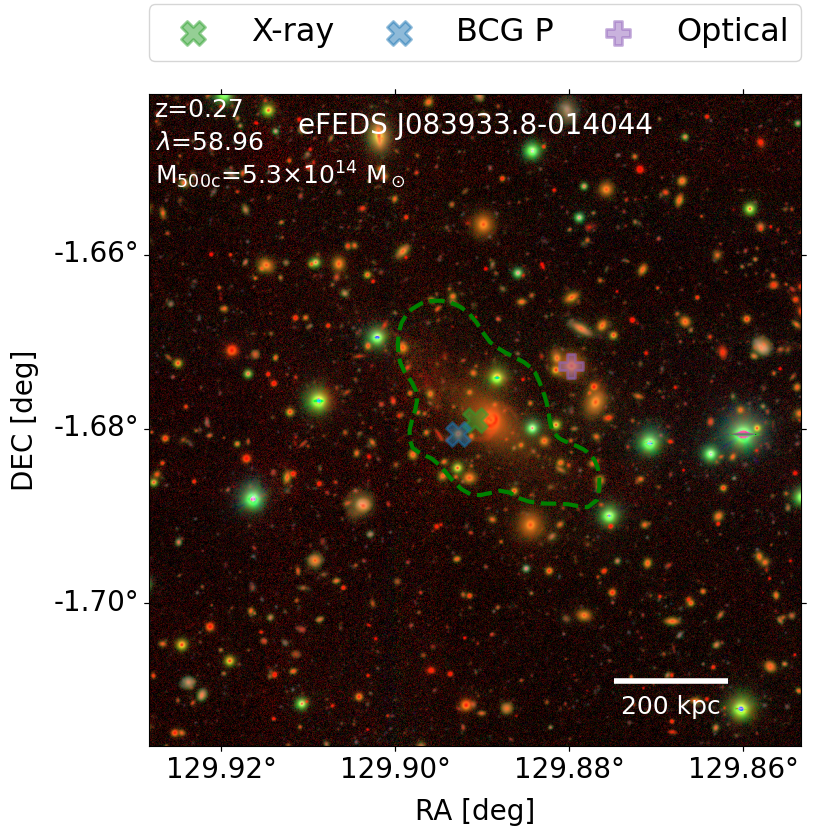}
    \includegraphics[width=0.73\columnwidth]{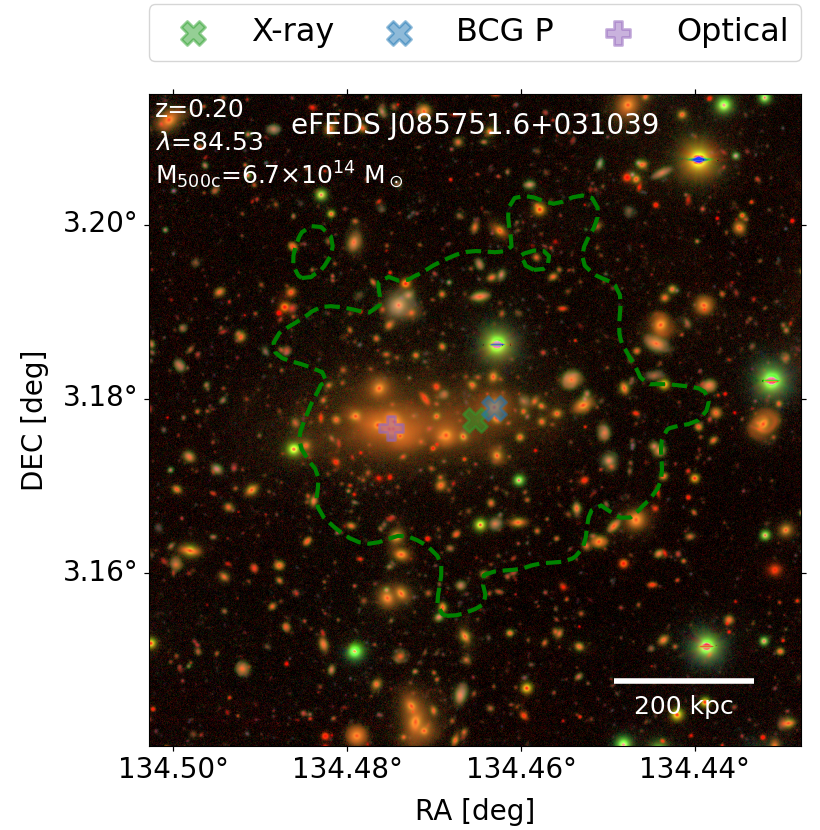}
    \includegraphics[width=0.73\columnwidth]{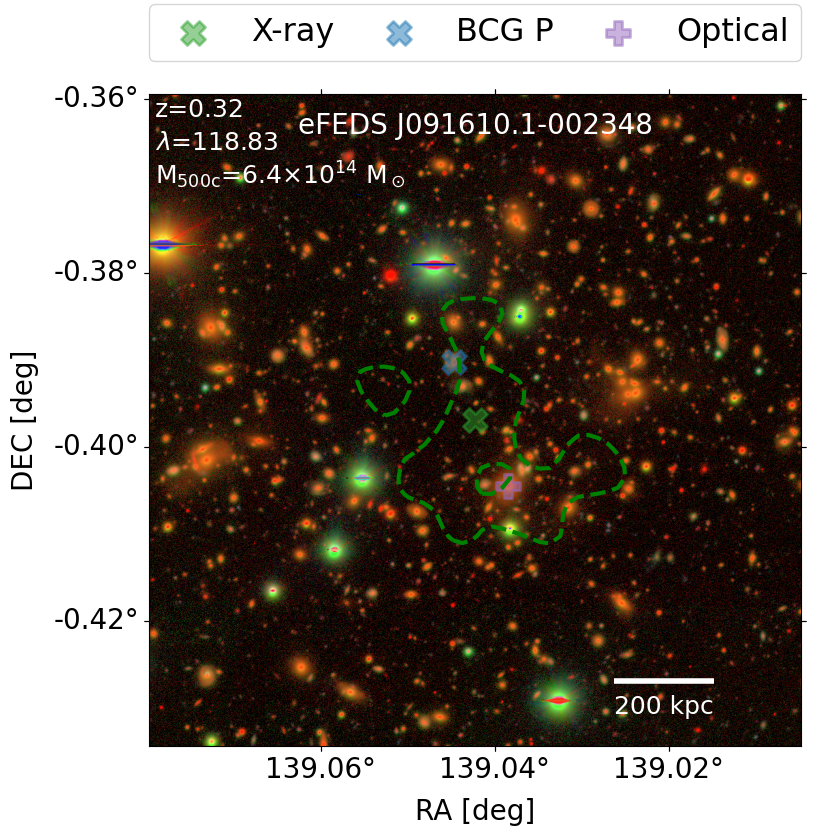}
    \includegraphics[width=0.73\columnwidth]{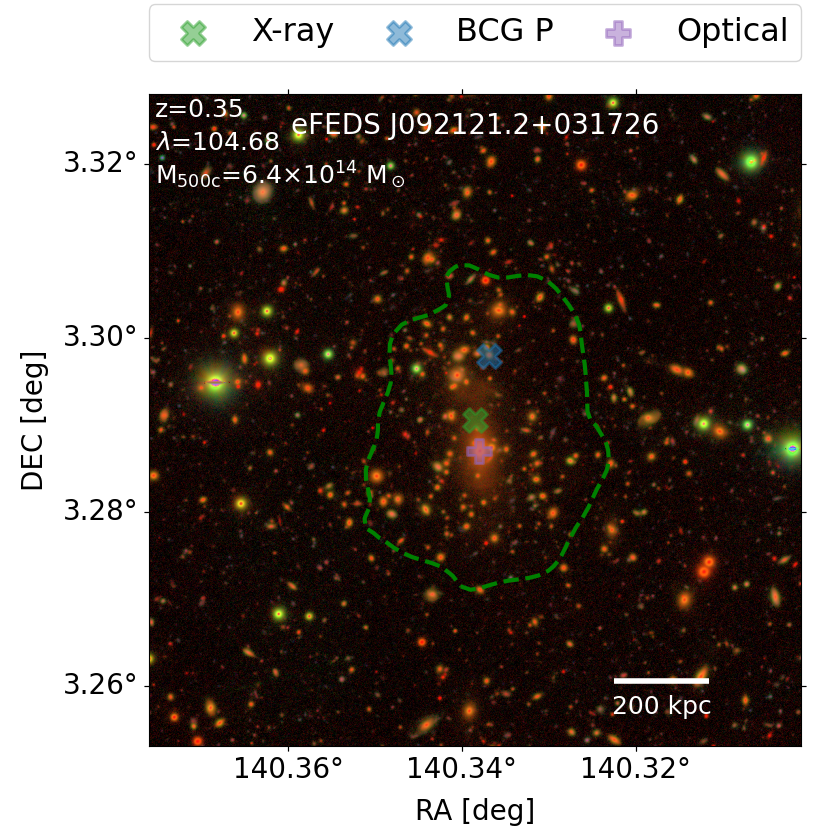}
    \includegraphics[width=0.73\columnwidth]{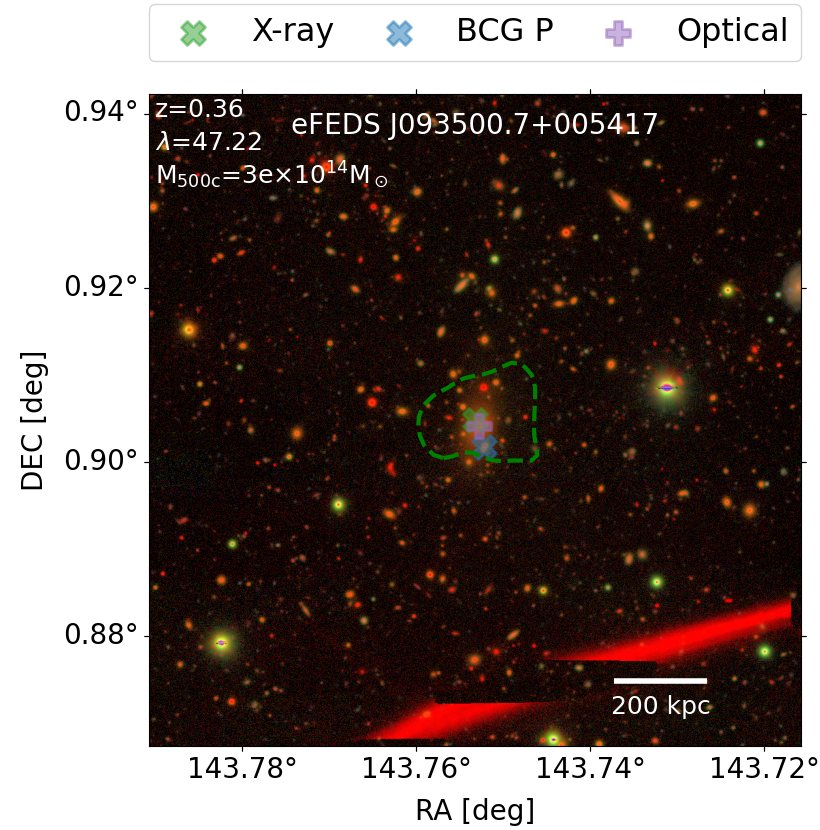}
    \includegraphics[width=0.73\columnwidth]{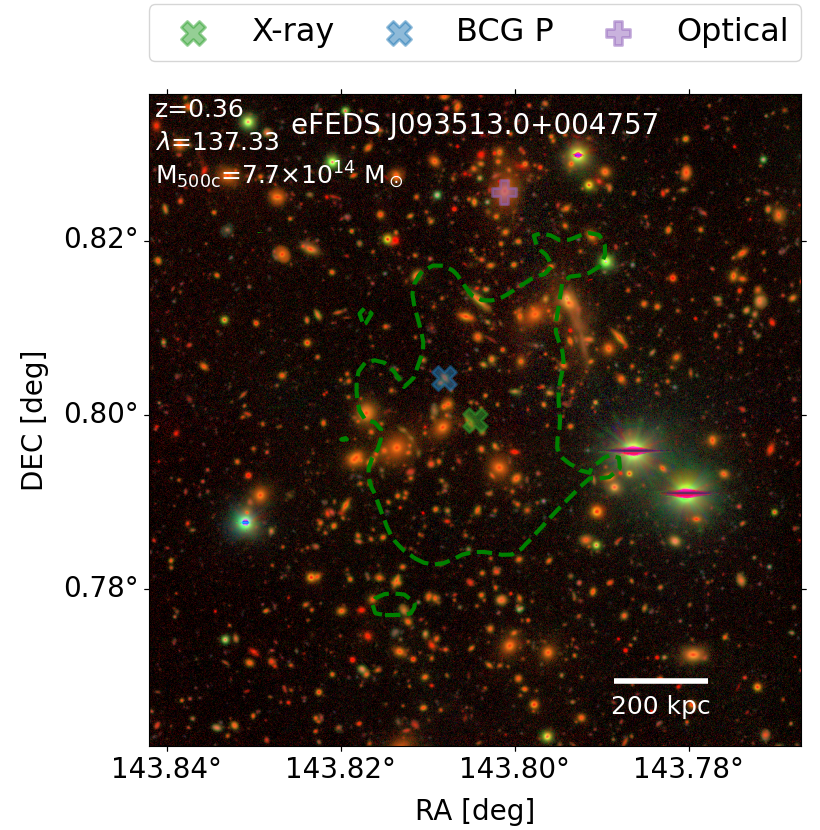}    
	\caption{eFEDS clusters. Each panel shows a different cluster with the relative name, redshift (z), richness ($\lambda$), and mass (M$_{\rm 500c}$). The optical image is an RGB cube built from HSC data using g, r, and z bands. The green lines denote the 3$\sigma$ contours of the X-ray emission. The green cross identifies the X-ray center found by eSASS, the blue one denotes the position of the galaxy with the largest membership probability, and the pink plus sign locates the optical center identified the centering algorithm of redMaPPer.}
    \label{fig:cluster_images}
\end{figure*}

Figure \ref{fig:cluster_images} shows six eFEDS clusters. The panels show clusters as seen in the optical band by HSC, using g, r, and z bands. The green dashed lines denote the 3$\sigma$ contours of the X-ray emission. The green cross identifies the X-ray center from eSASS. Two definitions of the optical center are shown: the galaxy with the largest membership probability (blue cross), and the optical center identified by the centering algorithm of redMaPPer (pink plus sign). In some cases, different characterizations of the center are in agreement. The cluster in the bottom-left panel in Figure \ref{fig:cluster_images} is an example. In other cases, the offsets are larger than hundreds of kiloparsecs, especially when the X-ray morphology has a complex structure (central-left panel in Figure \ref{fig:cluster_images}).

\begin{figure*}[h]
    \centering
    \includegraphics[width=0.66\columnwidth]{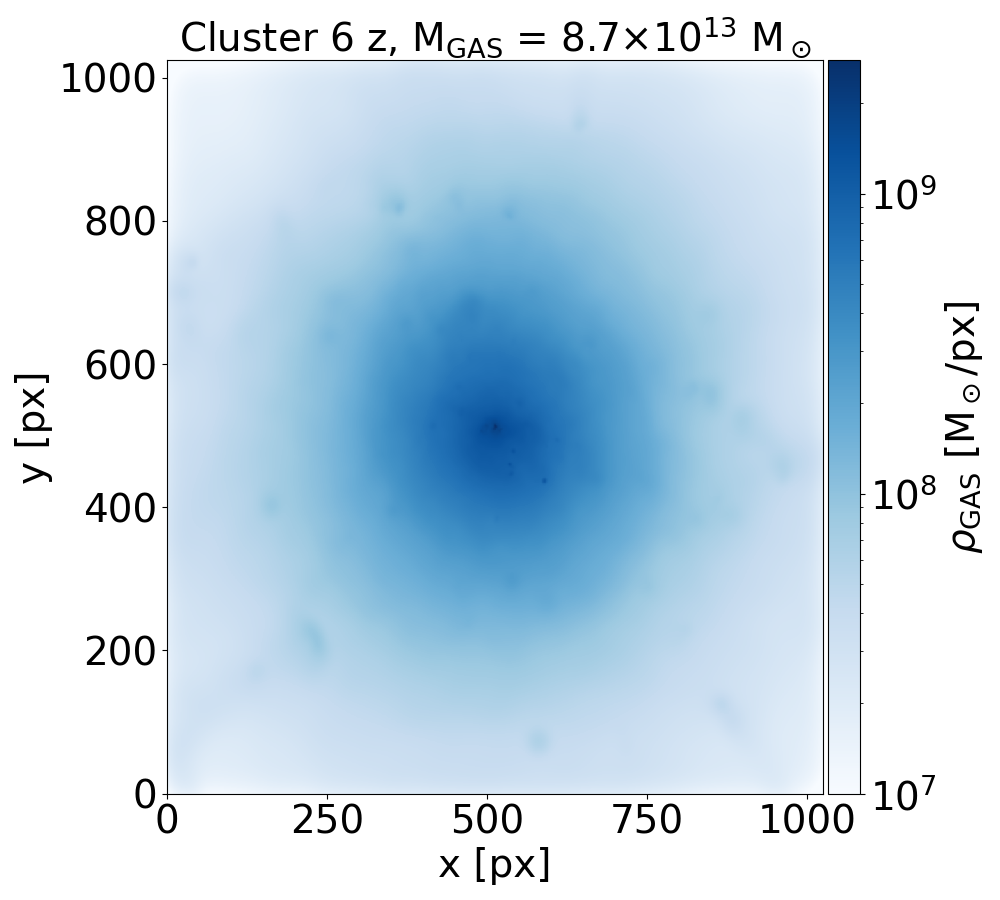}
    \includegraphics[width=0.66\columnwidth]{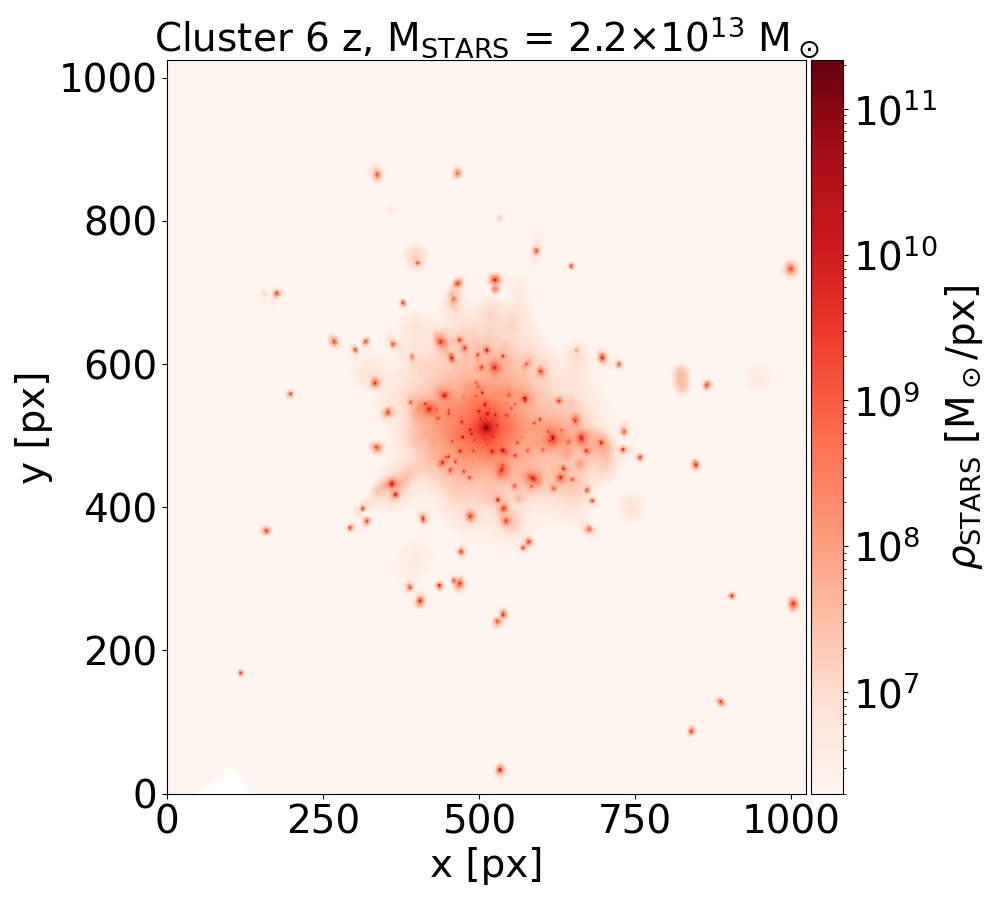}
    \includegraphics[width=0.66\columnwidth]{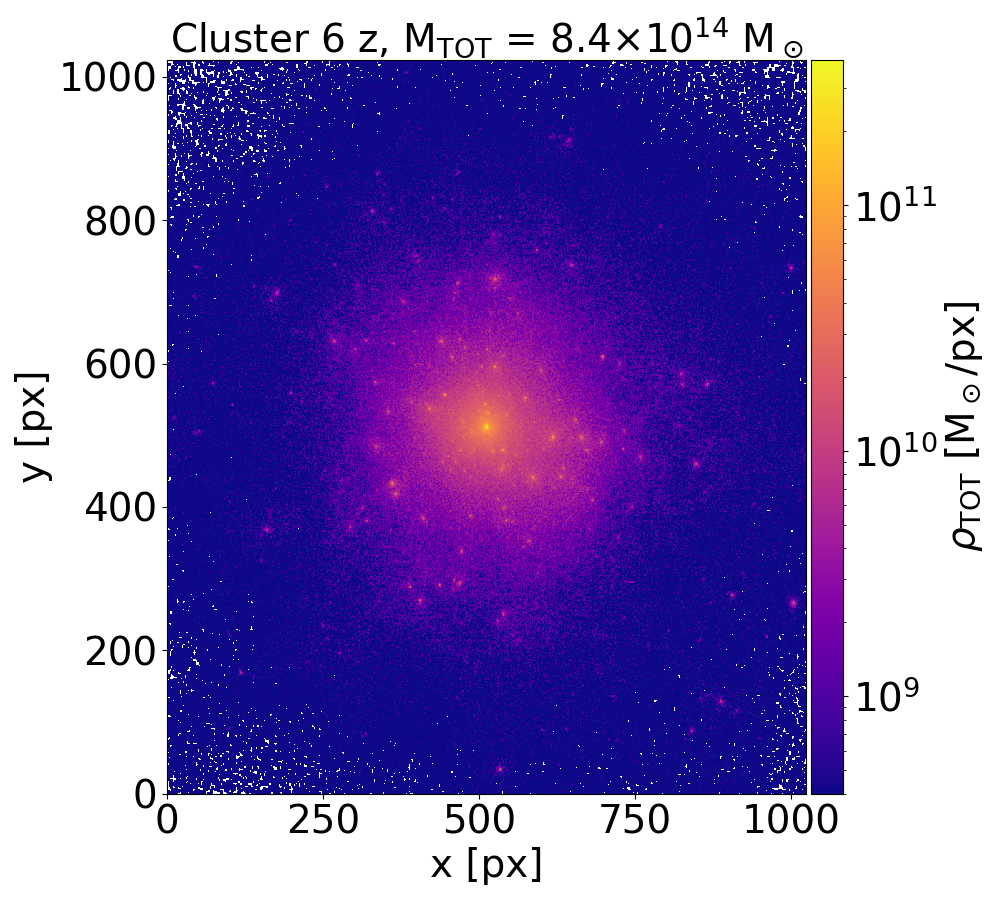}
    \includegraphics[width=0.66\columnwidth]{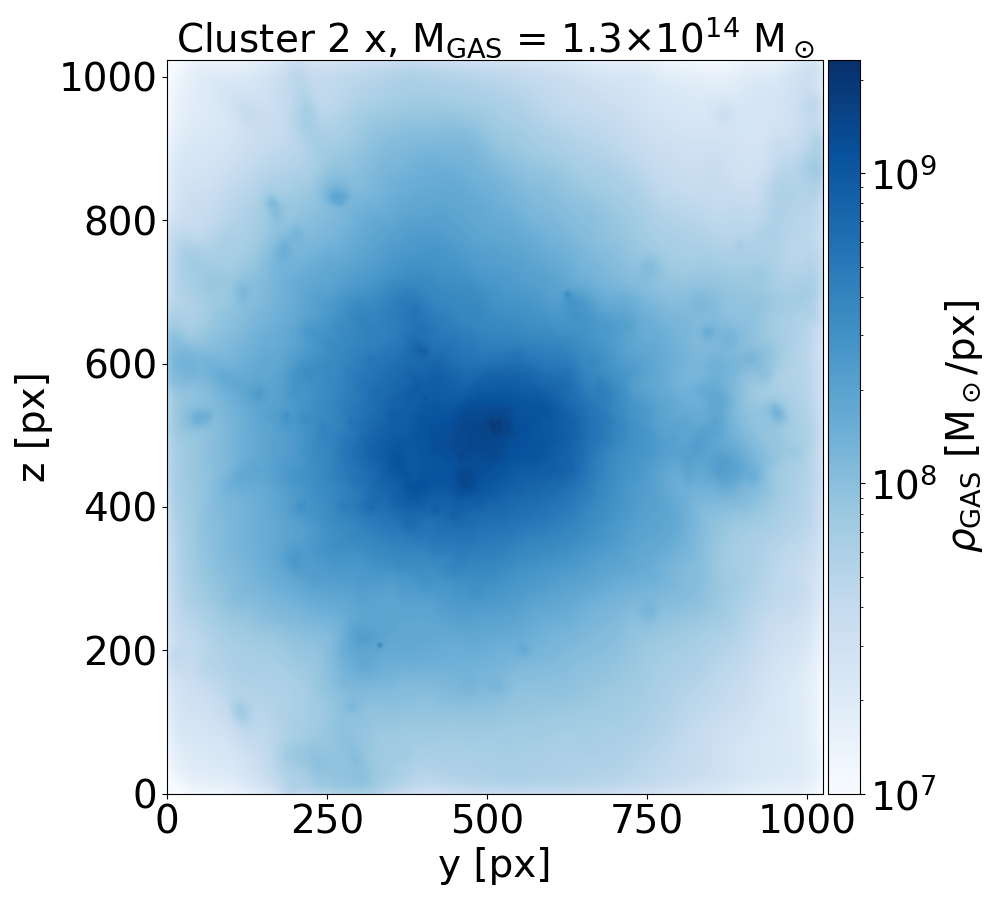}
    \includegraphics[width=0.66\columnwidth]{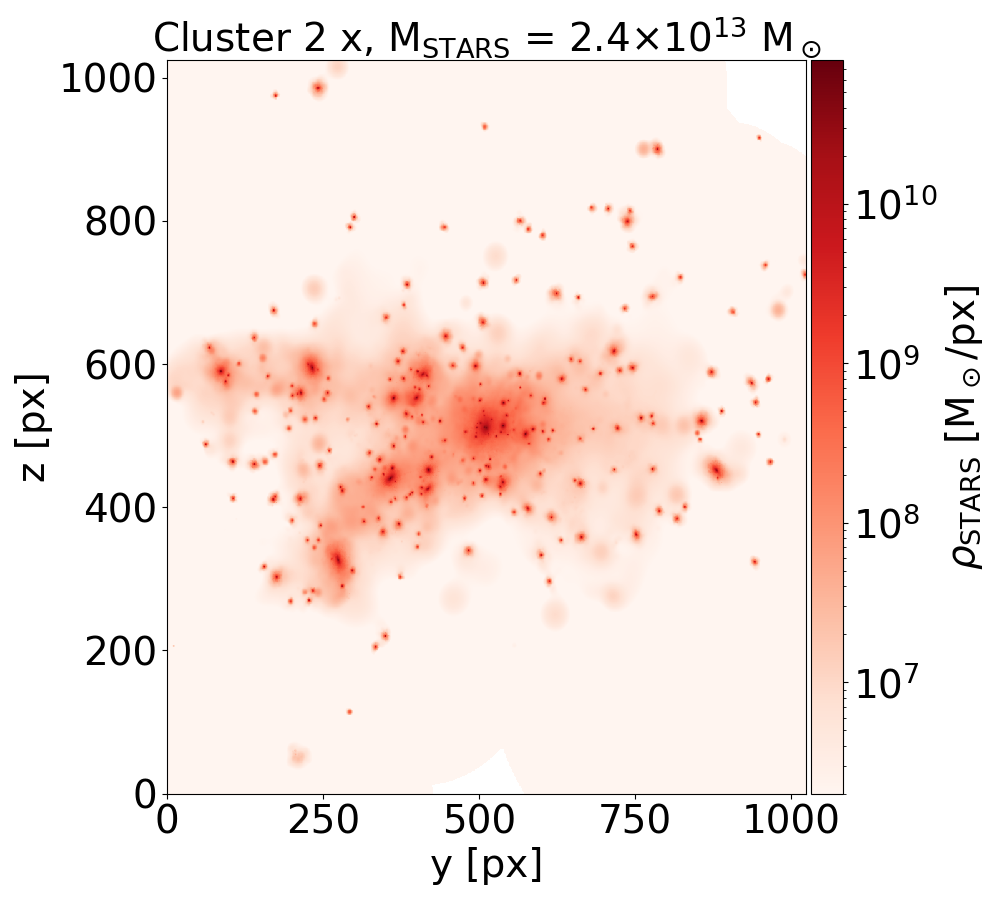}
    \includegraphics[width=0.66\columnwidth]{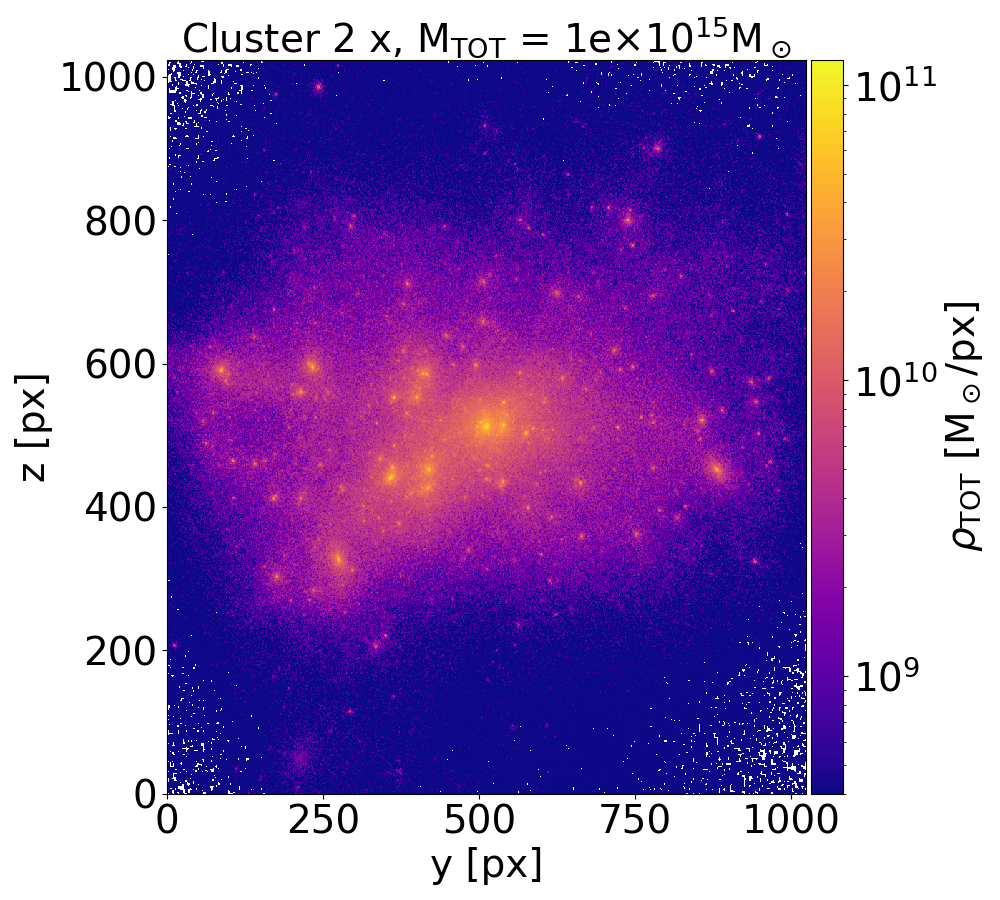} 
    \includegraphics[width=0.66\columnwidth]{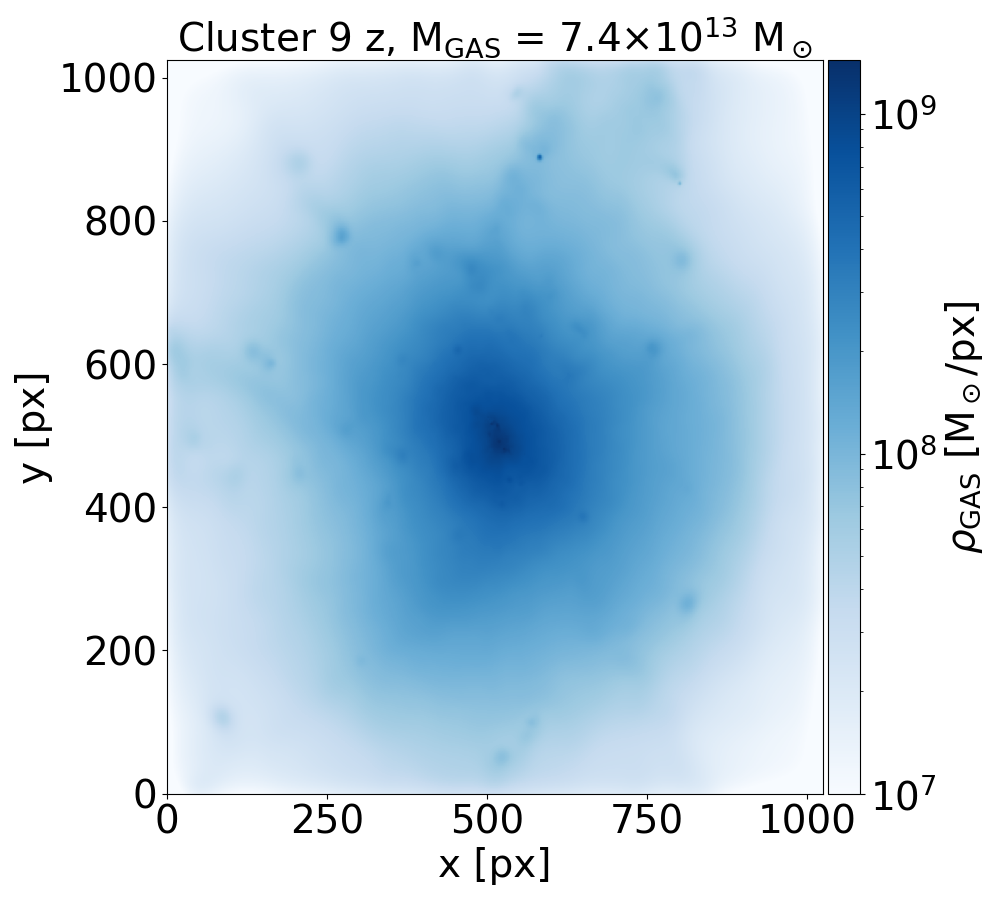}
    \includegraphics[width=0.66\columnwidth]{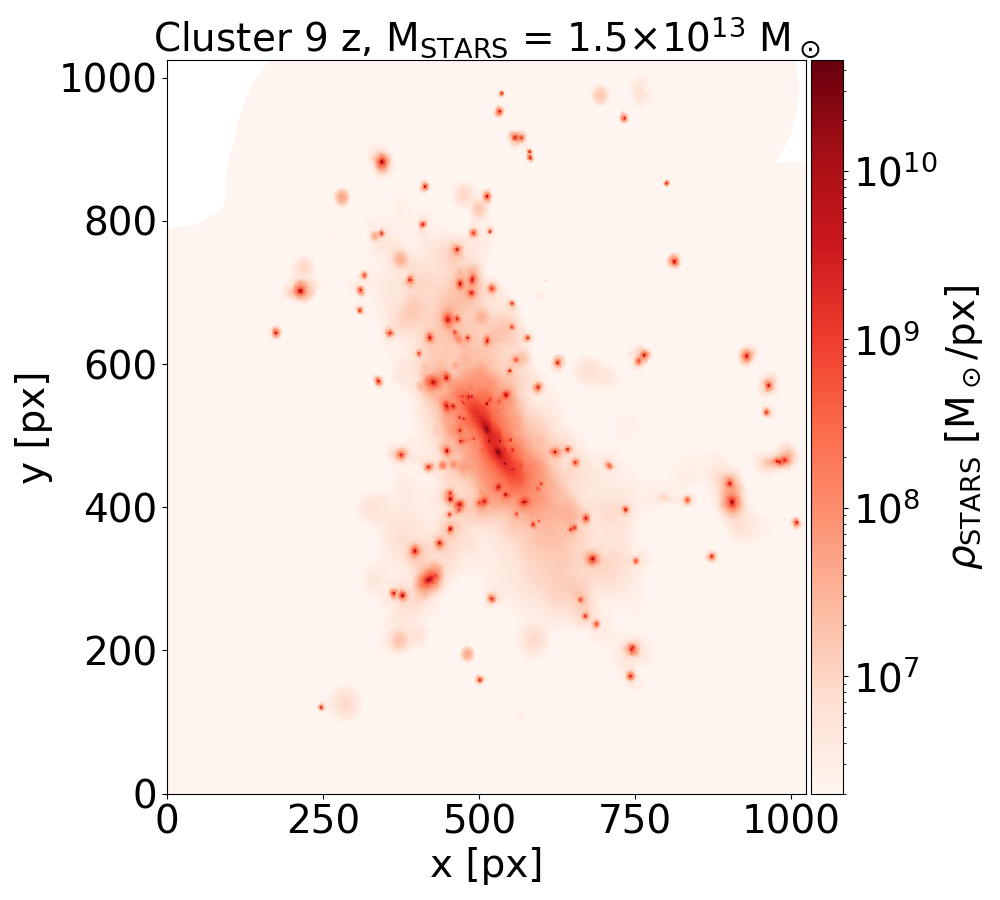}
    \includegraphics[width=0.66\columnwidth]{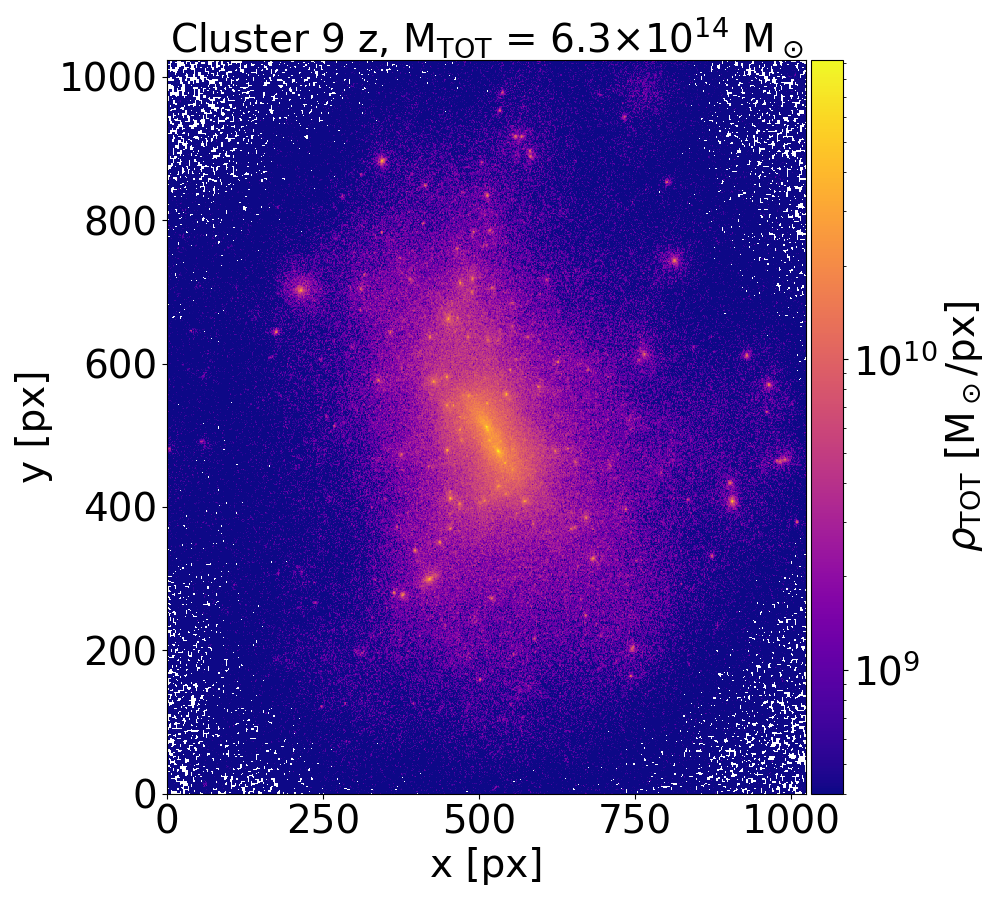}
	\caption{Clusters of galaxies in the Magneticum Box2/hr simulation at z=0.25. Each row displays one object. Each column shows a different component: the hot gas in the left-hand column, the stars in the middle one, and the total matter distribution in the right-hand one. The panels are 4$\times$R$_{\rm 500c}$ large. The side of the maps contains 1024 pixels, so the pixel size is $\sim$3.9$\times$10$^{-3}$R$_{\rm 500c}$. The figures are color coded according to the density of each component in units of solar masses per pixel. The first and third clusters are projected along the z axis, the middle one along the x axis. The mass of each component inside R$_{\rm 500c}$ is reported in the title of each panel.}
    \label{fig:cluster_images_magneticum}
\end{figure*}

Figure \ref{fig:cluster_images_magneticum} shows three of the most massive clusters in the snapshot at z=0.25 of the Magneticum simulation.
Each row displays one cluster. The left-hand column shows the hot gas component, the central one the stars, and the right-hand one the total matter distribution. The first cluster is relaxed: the distribution of the total matter is close to spherical and the hot gas is smooth. The cluster in the second row is in a transition phase, with a clear peak of the matter distribution in the center and some infalling satellites, that are also well traced by the star component. Finally, the third cluster is in the merging process: two main structures are colliding in the center, the stars are aligned along the merger direction and the gas distribution is more clumpy.

\bibliographystyle{aa}
\bibliography{biblio} 
\end{document}